\documentclass[%
 reprint,
 amsmath,amssymb,
 aps,
prb,
 superscriptaddress
]{revtex4-2}

\usepackage{graphicx}
\usepackage{dcolumn}
\usepackage{bm}
\usepackage{braket}
\newcommand{\1}{\mbox{1}\hspace{-0.25em}\mbox{l}}
\newcommand{\diff}{{\rm d}}
\usepackage{color}

\begin{document}

\preprint{APS/123-QED}

\title{Non-Bloch band theory in bosonic Bogoliubov--de Gennes systems}

\author{Kazuki Yokomizo}
\affiliation{Department of Physics, Tokyo Institute of Technology, 2-12-1 Ookayama, Meguro-ku, Tokyo, 152-8551, Japan}
\author{Shuichi Murakami}%
\affiliation{Department of Physics, Tokyo Institute of Technology, 2-12-1 Ookayama, Meguro-ku, Tokyo, 152-8551, Japan}
\affiliation{TIES, Tokyo Institute of Technology, 2-12-1 Ookayama, Meguro-ku, Tokyo, 152-8551, Japan}%





\begin{abstract}
In recent research, it has been shown that non-Hermitian systems exhibit sensitivity to boundaries, and it is caused by the non-Hermitian skin effect. In this work, we construct the non-Bloch band theory in bosonic Bogoliubov--de Gennes (BdG) systems. From our theory, we can calculate the generalized Brillouin zone and the energy spectrum in such systems with open boundary conditions in the thermodynamic limit, and we can thus discuss its non-Hermitian nature, despite Hermiticity of an original Hamiltonian. In fact, we find that the bosonic Kitaev-Majorana chain exhibits rich aspects of the non-Hermitian skin effect, such as instability against infinitesimal perturbations and reentrant behavior, in terms of the non-Bloch band theory. This result indicates that our theory is powerful tool for studying non-Hermitian nature in bosonic BdG systems.
\end{abstract}

\pacs{Valid PACS appear here}
\maketitle


%
%

\section{\label{sec1}Introduction}
In condensed matter physics, bosons describing low-energy excitations have been studied in various fields of physics. Such bosons include, for example, magnons, phonons, and photons. In many bosonic systems, a Hamiltonian is described in the form of the Bogoliubov--de Gennes (BdG) Hamiltonian. For example, in a magnon system, the dipolar interaction or the Dzyaloshinskii-Moriya interaction between spins leads to the bosonic BdG Hamiltonian with terms not preserving particle numbers~\cite{Zhang2013,Shindou2013,Shindou2013v2,Matsumoto2014,Zyuzin2016,Lu2018,Kondo2019,Joshi2019,Kondo2019v2,Hwang2020}.

In recent years, non-Hermitian physics of such bosonic BdG systems has been focused on and vigorously studied. For example, in a cold-atom system, the previous works~\cite{Fallani2004} experimentally observed the dynamical instability for the Bose-Einstein condensate caused by the contact interaction between bosons. From the theoretical perspective, the complex energy eigenvalues in this system lead to such instability~\cite{Nakamura2008}. In general cases, it was shown that when the bosonic BdG Hamiltonian is not positive definite, the energy eigenvalues of bosonic BdG systems become complex~\cite{Kawabata2019}. Thus non-Hermiticity emerges in many bosonic BdG systems.

In many previous works, non-Hermiticity in one-dimensional (1D) bosonic BdG systems has been investigated mainly under periodic boundary conditions. On the other hand, as long as we focus on a periodic chain, we cannot clarify non-Hermitian phenomena unique to such systems. For example, exponential growth of topological edge states occurs because these states can have complex energies, whereas bulk eigenstates have real energy eigenvalues~\cite{Barnett2013,Engelhardt2015,Furukawa2015,Galilo2015,Engelhardt2016}. Besides, in Ref.~[\onlinecite{McDonald2018}], it was theoretically proposed that in a bosonic BdG system, one can realize the non-Hermitian skin effect~\cite{Yao2018,Jin2019,Deng2019,Okuma2020,Borgnia2020,Yu2020,Yu2020v2,Lee2020,Kawabata2020,Yoshida2020,Zhang2020,Yi2020,Li2020,Kawabata2020v2,Okugawa2020,Brandenbourger2019,Gou2020,Xiao2020,Hofmann2020,Helbig2020} in which bulk states are localized at either edge of an open chain. These novel phenomena are caused by the existence of boundaries, and hence we should study non-Hermitian open chains. However, while some previous works studied non-Hermiticity in finite systems under open boundary conditions by numerical calculation, non-Hermitian physics in bosonic BdG systems in the thermodynamic limit has been unrevealed yet.

In our previous works~\cite{Yokomizo2019,Yokomizo2020,Yokomizo2020v2}, we established the non-Bloch band theory and studied non-Hermitian free-fermion systems with open boundary conditions in the limit of a large system size. In this work, we construct the non-Bloch band theory in bosonic BdG systems. By using our theory, one can investigate general non-Hermitian properties in bosonic BdG systems with open boundary conditions in the thermodynamic limit. In this study, we use the bosonic Kitaev-Majorana chain proposed in Ref.~[\onlinecite{McDonald2018}]. We find that this system exhibits rich aspects of the non-Hermitian skin effect, i.e., instability against infinitesimal perturbations and reentrant behavior. Interestingly, non-Hermiticity appears despite the Hermiticity of the original bosonic BdG Hamiltonian, and this behavior is well understood by the non-Bloch band theory.

This paper is organized as follows. In Sec.~\ref{sec2}, we review an eigenvalue problem of a bosonic BdG system in real space and construct the non-Bloch band theory in such a system. Based on this discussion, in Sec.~\ref{sec3}, we investigate non-Hermiticity and show a way to construct the Bogoliubov transformation in the bosonic Kitaev-Majorana chain. Finally we summarize the contents of this paper and discuss the applications of our theory to real physical systems in Sec.~\ref{sec4}.

%
%

\section{\label{sec2}Bosonic Bogoliubov--de Gennes systems}
In this section, we construct the non-Bloch band theory in bosonic systems described by the BdG Hamiltonian. To this end, we review the formulation of the eigenvalue problem of the bosonic BdG Hamiltonian in real space in the previous work.

%
%

\subsection{\label{sec2-1}Real-space Bogoliubov--de Gennes Hamiltonian}
First of all, we review the formulation of the eigenvalue problem of the bosonic BdG Hamiltonian in real space~\cite{Colpa1978,Shindou2013,Shindou2013v2,Matsumoto2014,Lieu2018}. We start with a Hermitian bosonic BdG Hamiltonian representing 1D bosonic tight-binding systems. The Hamiltonian in real space is written as
\begin{equation}
H=\frac{1}{2}\left({\bm a}^\dag~{\bm a}\right)H_{\rm BdG}\left( \begin{array}{c}
{\bm a}      \\
{\bm a}^\dag
\end{array}\right),
\label{eq1}
\end{equation}
where ${\bm a}=\left(\dots,a_{1,1},\dots,a_{1,q},\dots,a_{L,1},\dots,a_{L,q},\dots\right)$. We note that a unit cell is composed of $q$ degrees of freedom, and $a_{j,\sigma}~\left(\sigma=1,\dots,q\right)$ represents a bosonic annihilation operator at the $j$th unit cell. Then the BdG Hamiltonian $H_{\rm BdG}$ preserves the particle-hole symmetry (PHS)~\cite{Kawabata2019} defined as
\begin{equation}
\tau_xH_{\rm BdG}^{\rm T}\tau_x^{-1}=H_{\rm BdG}.
\label{eq2}
\end{equation}
Here $\tau_x$ and $\tau_z$ are defined as
\begin{eqnarray}
\tau_x=\left( \begin{array}{cc}
O  & \1 \\
\1 & O
\end{array}\right),~\tau_z=\left( \begin{array}{cc}
\1 & O   \\
O  & -\1
\end{array}\right),
\label{eq3}
\end{eqnarray}
respectively, where $O$ and $\1$ are a zero matrix and an identity matrix with the dimension equal to that of the vector ${\bm a}$, respectively. Hence a matrix form of $H_{\rm BdG}$ is given by~\cite{Kawabata2019}
\begin{eqnarray}
H_{\rm BdG}=\left( \begin{array}{cc}
M           & \Delta \\
\Delta^\dag & M^{\rm T}
\end{array}\right).
\label{eq4}
\end{eqnarray}
In analogy with the fermionic BdG Hamiltonian representing Bardeen-Cooper-Schrieffer superconductors, the Hermitian matrix $M=M^\dag$ represents the normal parts, and the symmetric matrix $\Delta=\Delta^{\rm T}$ does pairing terms. In order to get the energy eigenvalues of the systems, we should diagonalize $H_{\rm BdG}$ by basis transformation $\left({\bm a}^\dag~{\bm a}\right)=\left({\bm\alpha}^\dag~{\bm\alpha}\right)T^\dag$, where ${\bm\alpha}$ is another set of bosonic annihilation operators. Here, since ${\bm a}$ and ${\bm\alpha}$ satisfy bosonic commutation relations, $T$ must be a paraunitary matrix~\cite{Colpa1978,Shindou2013,Shindou2013v2,Matsumoto2014,Lieu2018} defined as
\begin{equation}
T^\dag\tau_zT=\tau_z.
\label{eq5}
\end{equation}

Here we can explicitly write the eigenvalue equation of the BdG Hamiltonian $\tau_zH_{\rm BdG}$ as
\begin{eqnarray}
\left(\tau_zH_{\rm BdG}\right)\left( \begin{array}{cc}
U & V^\ast \\
V & U^\ast
\end{array}\right)=\left( \begin{array}{cc}
U & V^\ast \\
V & U^\ast
\end{array}\right)\left( \begin{array}{cc}
\Lambda & O        \\
O       & -\Lambda
\end{array}\right), \nonumber\\
\label{eq6}
\end{eqnarray}
where $\Lambda$ is a diagonal matrix, and
\begin{eqnarray}
T=\left( \begin{array}{cc}
U & V^\ast \\
V & U^\ast
\end{array}\right)
\label{eq7}
\end{eqnarray}
is a paraunitary matrix. This indicates that since $\tau_z H_{\rm BdG}$ is non-Hermitian, some features of non-Hermitian physics may arise, although the original bosonic BdG Hamiltonian is Hermitian. We note that if $H_{\rm BdG}$ is positive definite, all the eigenenergies of the systems become real, and the systems can be regarded as Hermitian systems~\cite{Kawabata2019}. On the other hand, if not, the bosonic BdG systems are essentially non-Hermitian, and the eigenenergies are complex in general. Importantly, in this case, we can get the energy eigenvalue in an open chain by applying the non-Bloch band theory to such systems, as we will discuss in this paper.

%
%

\subsection{\label{sec2-2}Non-Bloch band theory}
Now we construct the non-Bloch band theory in 1D bosonic BdG systems with open boundary conditions. In the following, we assume that the ranges of the hopping in the normal terms $M$ and the pairing terms $\Delta$ are up to $N_s$ and $N_p$ unit cells, respectively. Then Eq.~(\ref{eq6}) can be explicitly written as
\begin{widetext}
\begin{eqnarray}
\left\{ \begin{array}{l}
\displaystyle\sum_j\sum_{\tau=1}^q\left[\sum_{i=-N_s}^{N_s}M_{i,\sigma\tau}u_{j+i,\tau}^\kappa+\sum_{i=-N_p}^{N_p}\Delta_{i,\sigma\tau}v_{j+i,\tau}^\kappa\right]=E^\kappa u_{j,\sigma}^\kappa \vspace{8pt}\\
\displaystyle\sum_j\sum_{\tau=1}^q\left[-\sum_{i=-N_p}^{N_p}\Delta_{-i,\tau\sigma}^\ast u_{j+i,\tau}^\kappa-\sum_{i=-N_s}^{N_s}M_{-i,\tau\sigma}v_{j+i,\tau}^\kappa\right]=E^\kappa v_{j,\sigma}^\kappa
\end{array}\right.~\left(\sigma=1,\dots,q\right),
\label{eq8}
\end{eqnarray}
\end{widetext}
where $u_{j,\sigma}^\kappa$ and $v_{j,\sigma}^\kappa$ are the $(\left(j,\sigma\right),\kappa)$ components of the matrices $U$ and $V$, respectively, and $M_{i,\sigma\tau}$ and $\Delta_{i,\sigma\tau}$ are the $(\sigma,\tau)$ components of the matrices $M$ and $\Delta$ representing the hopping to the $\left(-i\right)$th nearest unit cell. Here, thanks to spatial periodicity in the bulk, the eigenvectors $\left(u_{j,\sigma}^\kappa,v_{j,\sigma}^\kappa\right)$ can be given by the linear combination as
\begin{equation}
\left(u_{j,\sigma}^\kappa,v_{j,\sigma}^\kappa\right)=\sum_{m=1}^{4N}\left(u_\sigma^{\kappa,m},v_\sigma^{\kappa,m}\right)\left(\beta_m\right)^j~\left(\sigma=1,\dots,q\right),
\label{eq9}
\end{equation}
where $N=q\max\left(N_s,N_p\right)$, and $\beta=\beta_m$ are the solutions of the characteristic equation
\begin{equation}
\det\left[s_z{\cal H}_{\rm BdG}\left(\beta\right)-E\right]=0
\label{eq10}
\end{equation}
of the non-Bloch BdG matrix
\begin{eqnarray}
\left[{\cal H}_{\rm BdG}\left(\beta\right)\right]_{\sigma\tau}=\left( \begin{array}{cc}
\displaystyle\sum_{i=-N_s}^{N_s}M_{i,\sigma\tau}\beta^i            & \displaystyle\sum_{i=-N_p}^{N_p}\Delta_{i,\sigma\tau}\beta^i \vspace{8pt}\\
\displaystyle\sum_{i=-N_p}^{N_p}\Delta_{-i,\tau\sigma}^\ast\beta^i & \displaystyle\sum_{i=-N_s}^{N_s}M_{-i,\tau\sigma}\beta^i
\end{array}\right), \nonumber\\
\label{eq11}
\end{eqnarray}
where $s_z$ is the $2q\times2q$ matrix expressed as $s_z={\rm diag}\left(\1,-\1\right)$. We note that the characteristic equation (\ref{eq10}) is an algebraic equation for $\beta$ of degree $4N$. From the non-Bloch band theory~\cite{Yokomizo2019,Yokomizo2020,Yokomizo2020v2}, we show that the continuum energy band in a long open chain can be obtained from the condition
\begin{equation}
\left|\beta_{2N}\right|=\left|\beta_{2N+1}\right|,
\label{eq12}
\end{equation}
where the solutions of Eq.~(\ref{eq10}) are ordered in the following way:
\begin{equation}
\left|\beta_1\right|\leq\left|\beta_2\right|\leq\cdots\leq\left|\beta_{4N-1}\right|\leq\left|\beta_{4N}\right|.
\label{eq13}
\end{equation}
Then, according to the non-Bloch band theory, the trajectories of $\beta_{2N}$ and $\beta_{2N+1}$ on the complex plane give the generalized Brillouin zone (GBZ) for the complex wave number $\beta={\rm e}^{ik},~k\in{\mathbb C}$. In particular, one can show $\left|\beta_{2N}\right|=\left|\beta_{2N+1}\right|=1$ in Hermitian systems, leading to the conventional Brillouin zone, $k\in{\mathbb R}$. We note that in free-fermion systems, the previous works~\cite{Yokomizo2019,Kawabata2020,Zhang2020,Yokomizo2020,Yokomizo2020v2,Yi2020,Yang2020} found various features of the GBZ unique to non-Hermitian systems. We expect that the bosonic BdG systems inherit such features of the GBZ.

Thus Eq.~(\ref{eq9}) indicates that the paraunitary matrix $T$ can be given by the linear combination of the terms $\left(\beta_m\right)^j~\left(m=1,\dots,4N\right)$. In fact, since from the basis transformation $\left({\bm a}^\dag~{\bm a}\right)=\left({\bm\alpha}^\dag~{\bm\alpha}\right)T^\dag$, we can get
\begin{equation}
{\bm\alpha}^\dag={\bm a}^\dag U-{\bm a}V,
\label{eq14}
\end{equation}
the quasiparticle-creation operator $\alpha^\dag_{\kappa,\sigma}$ can be explicitly written as
\begin{eqnarray}
\alpha_{\kappa,\sigma}^\dag&=&\sum_j\left(u_{j,\sigma}^\kappa a_{j,\sigma}^\dag-v_{j,\sigma}^\kappa a_{j,\sigma}\right) \nonumber\\
&=&\sum_j\sum_{m=1}^{4N}\left[u_\sigma^{\kappa,m}\left(\beta_m\right)^ja_{j,\sigma}^\dag-v_\sigma^{\kappa,m}\left(\beta_m\right)^ja_{j,\sigma}\right] \nonumber\\
\label{eq15}
\end{eqnarray}
for $\sigma=1,\dots,q$. It is important that the coefficients $u_\sigma^{\kappa,m}$ and $v_\sigma^{\kappa,m}$ are determined by the open boundary conditions given as
\begin{equation}
\alpha_{\kappa,\sigma}\left(j=0\right)=\alpha_{\kappa,\sigma}\left(j=L+1\right)=0
\label{eq16}
\end{equation}
and by the condition that the operators $\alpha_{\kappa,\sigma}$ and $\alpha_{\kappa,\sigma}^\dag$ satisfy the boson statistics given as
\begin{equation}
\left[\alpha_{\kappa,\sigma},\alpha_{\kappa^\prime,\tau}^\dag\right]=\delta_{\kappa,\kappa^\prime}\delta_{\sigma,\tau}.
\label{eq17}
\end{equation}
Therefore, from Eqs.~(\ref{eq15})--(\ref{eq17}), we can get the Bogoliubov transformation diagonalizing the BdG Hamiltonian in terms of the non-Bloch band theory.

%
%

\section{\label{sec3}Bosonic Kitaev-Majorana chain}
In this section, we investigate the bosonic Kitaev-Majorana chain proposed in Ref.~[\onlinecite{McDonald2018}]. While the occurrence of the non-Hermitian skin effect in this system was proposed in the previous work, we show that the non-Hermitian skin effect is fragile against infinitesimal perturbations that couple between two skin modes. Furthermore, in a special case, we can derive the analytical representation of the Bogoliubov transformation from Eq.~(\ref{eq15}).

%
%

\subsection{\label{sec3-1}Non-Hermitian property}
First of all, we start with the real-space Hamiltonian of the bosonic Kitaev-Majorana chain. It is given by
\begin{eqnarray}
H&=&\sum_j\left[\mu a_j^\dag a_j+\frac{t}{2}\left(e^{i\phi}a_{j+1}^\dag a_j+e^{-i\phi}a_j^\dag a_{j+1}\right)\right. \nonumber\\
&&\left.+\frac{i\Delta}{2}\left(a_{j+1}^\dag a_j^\dag-a_ja_{j+1}\right)\right],
\label{eq18}
\end{eqnarray}
where $a_j$ is a bosonic annihilation operator at the $j$th site, and all the parameters are set to be positive real numbers, for simplicity. This model corresponds to the case of $q=N=1$ in Sec.~\ref{sec2}. Remarkably, although Hamiltonian (\ref{eq18}) is Hermitian, it exhibits non-Hermitian physics as we will discuss later. We note that the previous work only studied non-Hermitian properties in this model with the case of $\mu=0$ and $\phi=\pi/2$. In the following, when $\phi=0$ and $\phi=\pi/2$, we only focus on the case of $t>\Delta$ because the system is dynamically unstable if $t<\Delta$~\cite{McDonald2018}.

The non-Bloch BdG matrix for Eq.~(\ref{eq18}) is written as
\begin{eqnarray}
{\cal H}_{\rm BdG}\left(\beta\right)&=&\left[\mu+\frac{t}{2}\cos\phi\left(\beta+\beta^{-1}\right)\right]\sigma_0-\frac{\Delta}{2}\left(\beta+\beta^{-1}\right)\sigma_y \nonumber\\
&&-\frac{it}{2}\sin\phi\left(\beta-\beta^{-1}\right)\sigma_z,
\label{eq19}
\end{eqnarray}
where $\beta=e^{ik},~k\in{\mathbb C}$, $\sigma_0$ is a $2\times2$ identity matrix, and $\sigma_i~\left(i=x,y,z\right)$ are the Pauli matrices. As mentioned in Sec.~\ref{sec2}, we can get the GBZ and the continuum energy bands by the characteristic equation
\begin{eqnarray}
0&=&\det\left[\sigma_z{\cal H}_{\rm BdG}\left(\beta\right)-E\right] \nonumber\\
&=&\frac{1}{4}\left(\Delta^2-t^2\right)\left(\beta^2+\beta^{-2}\right)-\mu^2+E^2+\frac{1}{2}\left(\Delta^2-t^2\cos2\phi\right) \nonumber\\
&&-\left(\mu t\cos\phi-itE\sin\phi\right)\beta-\left(\mu t\cos\phi+itE\sin\phi\right)\beta^{-1} \nonumber\\
\label{eq20}
\end{eqnarray}
by applying condition (\ref{eq12}), i.e., $\left|\beta_2\right|=\left|\beta_3\right|$. The resulting GBZ is not a unit circle in the complex $\beta$ plane in general, leading to non-Hermitian physics in this system. Importantly, since this system is intrinsically non-Hermitian, it exhibits the non-Hermitian skin effect.

When $\mu=0$, it is shown that $\sigma_z{\cal H}_{\rm BdG}\left(\beta\right)$ can be transformed into a block-diagonal matrix form as
\begin{eqnarray}
P\left(\sigma_z{\cal H}_{\rm BdG}\left(\beta\right)\right)P^{-1}=\left( \begin{array}{cc}
{\cal H}_+\left(\beta\right) & O                            \\
O                            & {\cal H}_-\left(\beta\right)
\end{array}\right)
\label{eq21}
\end{eqnarray}
by the similarity transformation given by
\begin{eqnarray}
P=\frac{1}{\sqrt{2C\left(C+t\cos\phi\right)}}\left( \begin{array}{cc}
C+t\cos\phi & -i\Delta    \\
i\Delta     & C+t\cos\phi
\end{array}\right), \nonumber\\
\label{eq22}
\end{eqnarray}
where $C=\sqrt{t^2\cos^2\phi-\Delta^2}$. Then the diagonal elements of Eq.~(\ref{eq21}) are given by
\begin{equation}
{\cal H}_\pm\left(\beta\right)=\frac{t}{2i}\sin\phi\left(\beta-\beta^{-1}\right)\pm\frac{1}{2}\sqrt{t^2\cos^2\phi-\Delta^2}\left(\beta+\beta^{-1}\right).
\label{eq23}
\end{equation}
In other words, Eq.~(\ref{eq20}) can be factorized into two irreducible algebraic equations for $\beta$ and $E$. Therefore, in this case, this two-band model is decoupled into two systems described by ${\cal H}_\pm\left(\beta\right)$, which are Hatano-Nelson models without disorder~\cite{Hatano1996}. We note that the condition for continuum energy bands is also split for each band. Namely, they are given by $\left|\beta_1^\pm\right|=\left|\beta_2^\pm\right|$ for each band, where $\beta_{1,2}^\pm$ are the solutions of the equation $\det\left[{\cal H}_\pm\left(\beta\right)-E\right]=0$, respectively.

Before going to the discussion of non-Hermitian properties in this system, let us introduce the energy winding number~\cite{Okuma2020,Zhang2020}, which tells us the topological origin of the non-Hermitian skin effect. This topological invariant in a non-Hermitian system with periodic boundary conditions can be defined as
\begin{equation}
W\left(E\right)=\frac{1}{2\pi i}\int_0^{2\pi}\diff k\frac{\diff}{\diff k}\log\det\left[H\left(k\right)-E\right],
\label{eq24}
\end{equation}
where $k$ is the conventional Bloch wave number, and $H\left(k\right)$ is the conventional Bloch Hamiltonian written as $H\left(k\right)\equiv{\cal H}\left(\beta=e^{ik}\right),~k\in{\mathbb R}$ from our non-Bloch matrix ${\cal H}\left(\beta\right)$. For a given value of the reference energy $E\in{\mathbb C}$, when the energy winding number $W\left(E\right)$ takes nonzero values, the non-Hermitian skin effect occurs as a manifestation of non-Hermitian topology, and furthermore, there exists a difference between energy eigenvalues in a periodic chain and those in an open chain. In other words, when these energy eigenvalues are identical, $W\left(E\right)=0$ for any reference energies, and the non-Hermitian skin effect does not occur. Thus, by using Eq.~(\ref{eq24}), we can investigate the behavior of non-Hermitian systems. In the following, we discuss the physics of the model for three different cases of the value of the parameter $\phi$.

%
%

\subsubsection{\label{sec3-1-1}Case I: $\phi=0$}
In the case of $\mu=0$, the energy winding number (\ref{eq24}) for each band is always zero because from Eq.~(\ref{eq23}), the energy eigenvalues take either of the real values $\pm\sqrt{t^2-\Delta^2}\cos k$ for $k\in{\mathbb R}$. Furthermore, Eq.~(\ref{eq24}) of the system is also zero in the case of $\mu\neq0$. Therefore, the non-Hermitian skin effect does not occur, and the resulting GBZ is always a unit circle, identical with Hermitian cases.

%
%

\subsubsection{\label{sec3-1-2}Case I\hspace{-.1em}I: $\phi=\pi/2$}
\begin{figure}[]
\includegraphics[width=8.5cm]{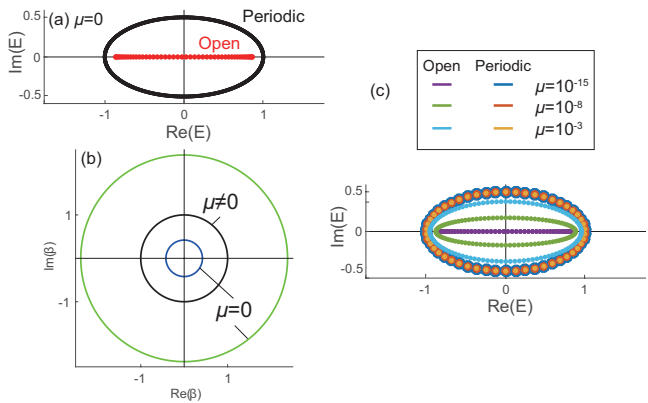}
\caption{\label{fig1}Eigenenergies and generalized Brillouin zone (GBZ) in the bosonic Kitaev-Majorana chain with various values of the parameter $\mu$ in the case of $\phi=\pi/2$. We set the parameters to be $t=1$ and $\Delta=0.7$. (a) Energy eigenvalues in an open chain (red) and in a periodic chain (black) with $\mu=0$. (b) Generalized Brillouin zone in the case of $\mu=0$ (green and blue) and of $\mu\neq0$ (black). The green (blue) circle describes the GBZ for the band with ${\cal H}_+\left(\beta\right)$ $\left[{\cal H}_-\left(\beta\right)\right]$ and has the radius $\left|\beta^+\right|$ ($\left|\beta^-\right|$). (c) Energy levels in a finite periodic chain and in a finite open chain with $\mu=10^{-15},10^{-8}$, and $10^{-3}$. We set the system size to be $L=50$. In the case of $L=50$, the critical value can be obtained as $\mu_0\simeq10^{-12}$. We note that the energy levels for a periodic chain with $\mu=10^{-15},10^{-8}$, and $10^{-3}$ almost overlap in the figure.}
\end{figure}

First of all, we begin with the case of $\mu=0$. Then, from Eq.~(\ref{eq23}), the energy winding number (\ref{eq24}) for each band with ${\cal H}_\pm\left(\beta\right)$ is $\pm1$, respectively, for any values of the reference energy within the region surrounded by an ellipse $E=t\sin k+i\Delta\cos k,~k\in{\mathbb R}$, which is the energy eigenvalue for a periodic chain, shown in black in Fig.~\ref{fig1}(a). Therefore, the non-Hermitian skin effect occurs in an open chain. In fact, we can confirm that the energy eigenvalue in an open chain completely differs from that in a periodic chain as shown in Fig.~\ref{fig1}(a), and the GBZ for each band is not a unit circle but a circle with the radius
\begin{equation}
\left|\beta^+\right|=\sqrt{\frac{t+\Delta}{t-\Delta}},~\left|\beta^-\right|=\sqrt{\frac{t-\Delta}{t+\Delta}},
\label{eq25}
\end{equation}
for the bands with ${\cal H}_\pm\left(\beta\right)$, respectively, as shown in Fig.~\ref{fig1}(b). It means that the eigenstates from ${\cal H}_+\left(\beta\right)$ and those from ${\cal H}_-\left(\beta\right)$ are localized at the opposite ends of an open chain.

While the non-Hermitian skin effect appears when $\mu=0$, it becomes unstable against perturbations by the infinitesimal values of the parameter $\mu$~\cite{Okuma2019,Li2020}. When $\mu$ becomes nonzero, the matrix $\sigma_z{\cal H}_{\rm BdG}\left(\beta\right)$ cannot be written as the block-diagonal matrix form (\ref{eq21}), which means that the two localized eigenstates couple with each other. Namely, this perturbation couples the two bands with the energy winding number $W\left(E\right)$ equal to $\pm1$, leading to a trivial value of $W\left(E\right)$, and thus the non-Hermitian skin effect is immediately suppressed. As a result, the GBZ becomes a unit circle when $\mu\neq0$ as shown in Fig.~\ref{fig1}(b), which means that the Bloch wave number takes real values, and the energy spectrum in an open chain coincides with that in a periodic chain. The transition from the nontrivial phase to the trivial phase discontinuously occurs as $\mu$ becomes nonzero. On the other hand, in a finite open chain, the energy levels continuously approach those in a finite periodic chain as the value of $\mu$ continuously increases, with a rapid change around the critical value $\mu_0$, as shown in Fig.~\ref{fig1}(c). Here, from Eq.~(\ref{eq25}), the order of $\mu_0$ is expected to be~\cite{Okuma2019}
\begin{equation}
\mu_0/t\simeq{\cal O}\left(\left|\beta^-\right|^L\right),
\label{eq26}
\end{equation}
where $L$ is the system size. For example, when $t=1,\Delta=0.7$, and $L=50$ adopted in Fig.~\ref{fig1}(c), we get $\mu_0\simeq10^{-12}$, and indeed, one can confirm this value of $\mu_0$ in a finite open chain; the energy levels with $\mu=10^{-15}<\mu_0$ largely deviate from those with $\mu=10^{-8}$ and $10^{-3}$ exceeding $\mu_0$. Thus the non-Hermitian skin effect exhibits instability against infinitesimal values of perturbation by $\mu$.

%
%

\subsubsection{\label{sec3-1-3}Case I\hspace{-.1em}I\hspace{-.1em}I: $\phi\neq0,\pi/2$}
\begin{figure*}[]
\includegraphics[width=17.5cm]{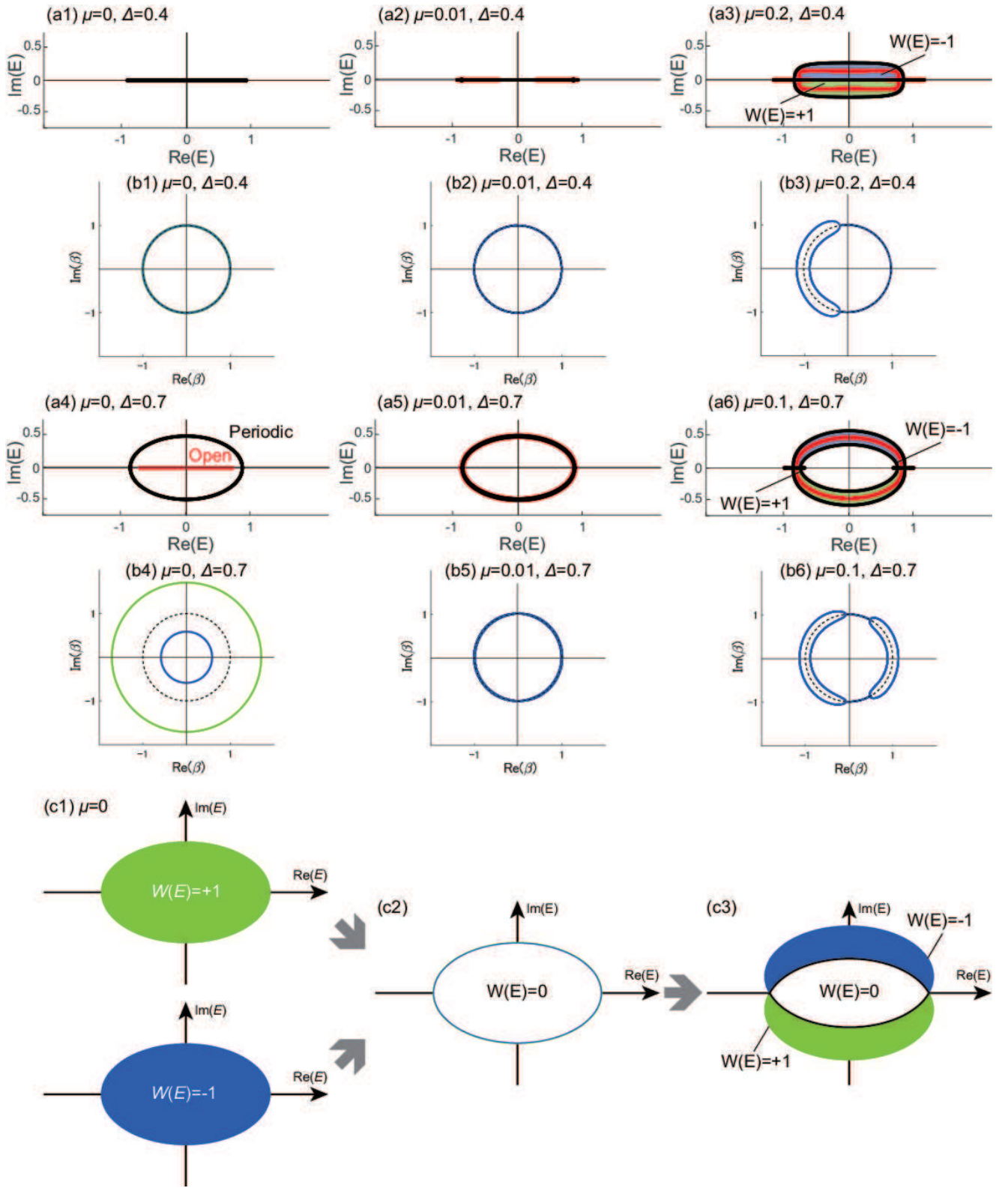}
\caption{\label{fig2}Eigenenergies and generalized Brillouin zone in the bosonic Kitaev-Majorana chain with various values of the parameter $\mu$ in the case of $\phi=\pi/3$. We set the parameters to be $t=1$. (a1--a6) Continuum energy bands in an open chain (red) and the energy spectrum in a periodic chain (black). In (a3) and (a6), the region in green (blue) represents that the energy winding number $W\left(E\right)$ takes $+1$ ($-1$). (b1--b6) Generalized Brillouin zone with the colored lines. The black broken line expresses a unit circle, meaning the conventional Brillouin zone. (c1--c3) Schematic figures of the reentrant behavior of the energy winding number $W\left(E\right)$ through an increase of the value of the parameter $\mu$. At (c1) $\mu=0$, the two bands with $W\left(E\right)=\pm1$ are decoupled. When we increase $\mu$, they couple with each other, leading to $W\left(E\right)=0$ as shown in (c2). A further increase of $\mu$ leads to nonzero values of $W\left(E\right)$ because of the deviation between two energy spectra for the two bands in a periodic chain as shown in (c3). The region in green (blue) on the complex energy plane describes the region with $W\left(E\right)$ being $1$ ($-1$).}
\end{figure*}
First of all, when $\mu=0$, whether the non-Hermitian skin effect occurs or not depends on the values of the system parameters. When $t\left|\cos\phi\right|>\Delta$, the non-Hermitian skin effect disappears because the energy winding number (\ref{eq24}) becomes zero for any reference energies. This is because the energy eigenvalues for $k\in{\mathbb R}$ take the real values $t\sin\phi\sin k\pm\sqrt{t^2\cos^2\phi-\Delta^2}\cos k$. On the other hand, when $t\left|\cos\phi\right|<\Delta$, the system exhibits the non-Hermitian skin effect. In fact, the above discussion is consistent with the numerical results as shown in Figs.~\ref{fig2}(a1) and \ref{fig2}(b1) and in Figs.~\ref{fig2}(a4) and \ref{fig2}(b4).

Next we focus on the case of $\mu\neq0$. When $\mu$ is a positive infinitesimal, $\mu=+0$, the GBZ is a unit circle, regardless of the value of $\Delta$. It is seen in Figs.~\ref{fig2}(b2) and \ref{fig2}(b5) for $\mu=0.01$, where we note that because $\mu$ is small but finite, the GBZ slightly deviates from a unit circle, but this deviation is tiny when $\mu\ll{\cal O}\left(t,\Delta\right)$. Accordingly, the eigenenergies in an open chain are almost identical with those in a periodic chain, shown in Figs.~\ref{fig2}(a2) and \ref{fig2}(a5), and the non-Hermitian skin effect is almost absent. As the value of $\mu$ increases, the deviation of the GBZ from a unit circle becomes prominent [Figs.~\ref{fig2}(b3) and \ref{fig2}(b6)]. Therefore, in particular, when $t\left|\cos\phi\right|<\Delta$, as the value of $\mu$ increases, the non-Hermitian skin effect disappears once and it reoccurs as shown in Figs.~\ref{fig2}(a4)--\ref{fig2}(a6) and (b4)--\ref{fig2}(b6). In this sense, we call this phenomenon reentrant behavior. It is noted that the reappearance of the deviation of the GBZ and that of the non-Hermitian skin effect occurs in $\mu>0$ as a crossover, and not a phase transition.

We note that the GBZ in Figs.~\ref{fig2}(b3) and \ref{fig2}(b6) partially overlaps a unit circle shown in a black broken line. This means that the Bloch wave number takes real values because the BdG Hamiltonian (\ref{eq18}) is positive definite in some regions on the complex $\beta$ plane. On the other hand, the positive definiteness of Eq.~(\ref{eq18}) is partially broken in the other regions, and the shape of the GBZ deviates from a unit circle in such regions. Thus the Hermiticity and the non-Hermiticity can coexist in a set of the system parameters.

Now we explain the reason for the reentrant behavior of the non-Hermitian skin effect upon changing the value of $\mu$ as shown in Figs.~\ref{fig2}(c1)--\ref{fig2}(c3). As mentioned above, when $\mu=0$, this system can be regarded as two decoupled Hatano-Nelson models [Fig.~\ref{fig2}(c1)]. In this case, for any values of the reference energy $E$ in the region surrounded by the energy band, the values of the energy winding number $W\left(E\right)$ are equal to $\pm1$ for the two decoupled bands, leading to one skin mode each. On the other hand, for the infinitesimal values of $\mu$, the two bands couple with each other, and the total energy winding number in this system is merely summed over the two bands to approximately become zero [Fig.~\ref{fig2}(c2))]. Therefore the non-Hermitian skin effect almost disappears, as we have already seen in Figs.~\ref{fig2}(a5) and \ref{fig2}(b5). This disappearance gradually occurs when the system size is not large, and it becomes sharp at $\mu$ being infinitesimal in the limit of a large system size. Furthermore, as the value of $\mu$ increases, the splitting of the two bands becomes significant, and the region with $W\left(E\right)\neq0$ on the complex energy plane appears due to the deviation of the two bands [Fig.~\ref{fig2}(c3)]. Therefore, these regions give rise to the two skin modes localized at both ends of an open chain. We note that in the case of $\phi=\pi/2$, this deviation does not occur because the two bands are accidentally degenerate for any values of $\mu$.

\begin{figure}[]
\includegraphics[width=8.5cm]{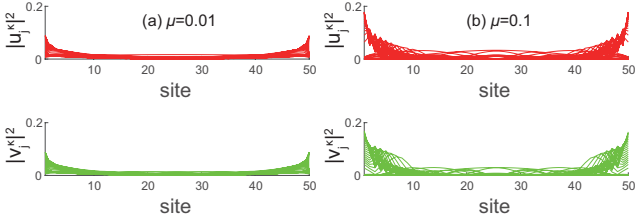}
\caption{\label{fig3} Real-space distribution of the coefficients $u_j^\kappa$ and $v_j^\kappa$ for $\kappa=1,\dots,50$ in the Bogoliubov transformation in a finite open chain with $t=1,\Delta=0.7,\phi=\pi/3$, and (a) $\mu=0.01$ and (b) $\mu=0.1$, respectively. We set the system size to be $L=50$.}
\end{figure}

Finally, in order to see how the non-Hermitian skin effect appears in real space, we show the numerical result of the real-space distribution of the coefficients $u_j^\kappa$ and $v_j^\kappa$ included in the Bogoliubov transformation with $t=1,\Delta=0.7,\phi=\pi/3$, and $\mu=0.01$ adopted in Figs.~\ref{fig2}(a5) and \ref{fig2}(b5), and $\mu=0.1$ adopted in Figs.~\ref{fig2}(a6) and \ref{fig2}(b6). When $\mu=0.01$, the non-Hermitian skin effect almost does not occur, and their distribution is almost uniform [Fig.~\ref{fig3}(a)]. In contrast, when $\mu=0.1$, the components of $u_j^\kappa$ and $v_j^\kappa$ at either end of an open chain have larger values than any other sites [Fig.~\ref{fig3}(b)]. This indicates that the non-Hermitian skin effect reoccurs as the values of $\mu$ increase from $\mu=0.01$, and therefore, this numerical result is consistent with the above discussion.

%
%

\subsection{\label{sec3-3}Analytical representation of Bogoliubov transformation}
In this section, we study the Bogoliubov transformation in a special case, i.e., $\mu=0$ and $\phi=\pi/2$. For simplicity, we assume $t>\Delta$ in the following. In this case, from Eq.~(\ref{eq25}), we can get an analytical form of the Bogoliubov transformation in a finite open chain with the system size $L$. From Eq.~(\ref{eq23}), the values of $\beta_m^\pm~(m=1,2)$ are explicitly written as
\begin{eqnarray}
\left\{ \begin{array}{l}
\displaystyle\beta_1^+=i\sqrt{\frac{t+\Delta}{t-\Delta}}e^{ik},~\beta_2^+=i\sqrt{\displaystyle\frac{t+\Delta}{t-\Delta}}e^{-ik}, \vspace{8pt}\\
\displaystyle\beta_1^-=i\sqrt{\frac{t-\Delta}{t+\Delta}}e^{ik},~\beta_2^-=i\sqrt{\displaystyle\frac{t-\Delta}{t+\Delta}}e^{-ik},
\end{array}\right.
\label{eq27}
\end{eqnarray}
where $k\in{\mathbb R}$. We note that theses complex Bloch wave numbers form the continuum energy band as $E=\sqrt{t^2-\Delta^2}\cos k$. Then we can determine the coefficients included in Eq.~(\ref{eq15}) so as to satisfy conditions (\ref{eq16}) and (\ref{eq17}) and can get the analytical representation of the quasiparticle-creation operator $\alpha_n^\dag$ as
\begin{eqnarray}
\alpha_n^\dag&=&\sum_{j=1}^L\sum_{m=1}^4\left[u^{\kappa,m}\left(\beta_m\right)^ja_j^\dag-v^{\kappa,m}\left(\beta_m\right)^ja_j\right] \nonumber\\
&=&\frac{1}{2i\sqrt{2\left(L+1\right)}} \nonumber\\
&&\times\sum_{j=1}^L\left\{\left[\left(\beta_1^+\right)^j-\left(\beta_2^+\right)^j+\left(\beta_1^-\right)^j-\left(\beta_2^-\right)^j\right]a_j\right. \nonumber\\
&&\left.-\left[\left(\beta_1^+\right)^j-\left(\beta_2^+\right)^j-\left(\beta_1^-\right)^j+\left(\beta_2^-\right)^j\right]a_j^\dag\right\} \nonumber\\
&=&\sqrt{\frac{2}{L+1}}\sum_{j=1}^Li^j\left[\sin\left(k_nj\right)\cosh\left(rj\right)a_j^\dag\right. \nonumber\\
&&\left.-\sin\left(k_nj\right)\sinh\left(rj\right)a_j\right],
\label{eq28}
\end{eqnarray}
where
\begin{equation}
e^{2r}=\frac{t+\Delta}{t-\Delta},
\label{eq29}
\end{equation}
and $k_n=n\pi/\left(L+1\right)~(n=1,\dots,L)$. We note that these results obtained here agree with those in the previous work~\cite{McDonald2018}. We can systematically get the Bogoliubov transformation in terms of the non-Bloch band theory in general cases. Thus the non-Bloch band theory proves to be a powerful tool in the study of bosonic BdG systems.

%
%

\section{\label{sec4}Summary}
In our work, we construct the non-Bloch band theory and study non-Hermitian nature in bosonic BdG systems. In such bosonic systems, although the original BdG Hamiltonian is Hermitian, the systems are intrinsically non-Hermitian. This is because the BdG Hamiltonian is locally a Hermitian operator but not a self-adjoint operator due to boundary conditions. In other words, since boundary conditions break the self-adjoint nature of the operator, eigenvalues of the BdG Hamiltonian can take complex values.

In fact, we analyze the non-Hermitian properties of the bosonic Kitaev-Majorana chain (\ref{eq18}) and find that this system exhibits rich aspects of the non-Hermitian skin effect with various values of the system parameters. Here we emphasize that one can analyze this non-Hermiticity only via the non-Bloch band theory. In this theory, in terms of the complex wave number $\beta=e^{ik},~k\in{\mathbb C}$, we can transform the matrix $\sigma_z{\cal H}_{\rm BdG}\left(\beta\right)$ into the block-diagonal form (\ref{eq21}) when $\mu=0$. Furthermore, based on this theory, interestingly, this model with $\phi\neq0,\pi/2$ exhibits the reentrant behavior of the non-Hermitian skin effect when $\mu\neq0$, which has not been found. It is a promising direction to observe this reentrant phenomenon in experiments, such as the Bose-Einstein condensate in cold-atom systems~\cite{Fallani2004,Anderson1995} and in magnon systems~\cite{Demokritov2006}.

We comment on a bulk-edge correspondence between a topological invariant and existence of topological edge states in bosonic BdG systems. Since the bosonic Kitaev-Majorana chain is classified as class A in Altland-Zirnbauer classes, it is topologically trivial, and it does not show the conventional bulk-edge correspondence.

When the BdG Hamiltonian is a positive-definite matrix, the system can be regarded as a Hermitian system, and the Hamiltonian can be diagonalized by using the Cholesky decomposition~\cite{Colpa1978}. It is exemplified by the bosonic Kitaev-Majorana chain with a sufficiently large value of the parameter $\mu$. In this case, energy spectra of an open chain and a periodic chain are asymptotically the same in the limit of the large system. On the other hand, if not, the system is essentially a non-Hermitian system, and the calculation method by using the Cholesky decomposition cannot be applied to this non-Hermitian system. In this sense, our theory is a powerful tool for studying non-Hermitian nature in bosonic BdG systems with open boundary conditions.

Finally, we discuss experimental applications of our theory established here. So far, many previous works proposed that various bosonic systems can be described by the BdG Hamiltonian, and in such systems, the non-Hermitian physics has intensively been studied, such as the dynamical instability in cold-atom systems~\cite{Barnett2013,Engelhardt2015,Furukawa2015,Galilo2015,Engelhardt2016}. Besides, the non-Hermitian nature of magnon systems has been also focused on in recent years~\cite{Kondo2020}. Hence, in these systems, we expect experimental observations of localization states induced by the non-Hermitian skin effect. Furthermore, through the observation of the skin modes, we can also confirm that the energy winding number (\ref{eq24}) takes nonzero values. We note that Eq.~(\ref{eq24}) cannot be associated with measurable quantities, such as transport coefficients. Therefore, the nonzero energy winding number can be indirectly observed through presence or absence of bulk skin modes and through observations of a complex energy spectrum in a periodic chain~\cite{Wang2021}.

Thus bosonic BdG systems are useful for investigating some non-Hermitian properties. In our work, we show that non-Hermitian phenomena in bosonic BdG systems are accessible via the non-Bloch band theory, and it paves a way to an implementation of the non-Hermitian phenomenon, such as the non-Hermitian skin effect, in many bosonic BdG systems.

%
%

\begin{acknowledgments}
We are grateful to Ryo Okugawa for valuable discussion. This work was supported by JSPS KAKENHI (Grant No.~JP18H03678) and by the MEXT Elements Strategy Initiative to Form Core Research Center (TIES) (Grant No.~JPMXP0112101001). K.Y. was also supported by JSPS KAKENHI Grant No.~JP18J22113.
\end{acknowledgments}

%
%

\appendix

%
%

\section{\label{secA}Particle-hole symmetry of bosonic Bogoliubov--de Gennes Hamiltonian}
In this Appendix, we prove that the BdG Hamiltonian $H_{\rm BdG}$ has the PHS defined in Eq.~(\ref{eq2}). We can rewrite the form of the real-space Hamiltonian (\ref{eq1}) as
\begin{eqnarray}
H&=&\frac{1}{2}\left({\bm a}~{\bm a}^\dag\right)\left(\tau_xH_{\rm BdG}\tau_x^{-1}\right)\left( \begin{array}{c}
{\bm a}^\dag \\
{\bm a}
\end{array}\right) \nonumber\\
&=&\frac{1}{2}\left({\bm a}^\dag~{\bm a}\right)\left(\tau_xH_{\rm BdG}\tau_x^{-1}\right)^{\rm T}\left( \begin{array}{c}
{\bm a}      \\
{\bm a}^\dag
\end{array}\right)-\frac{1}{2}{\rm Tr}\left(\tau_zH_{\rm BdG}\right) \nonumber\\
&=&\frac{1}{2}\left({\bm a}^\dag~{\bm a}\right)\left(\tau_xH_{\rm BdG}^{\rm T}\tau_x^{-1}\right)\left( \begin{array}{c}
{\bm a}      \\
{\bm a}^\dag
\end{array}\right)-\frac{1}{2}{\rm Tr}\left(\tau_zH_{\rm BdG}\right), \nonumber\\
\label{eqappA1}
\end{eqnarray}
where $\tau_x$ and $\tau_z$ are defined in Eq.~(\ref{eq3}). Since Eq.~(\ref{eqappA1}) should be equal to Eq.~(\ref{eq1}), we can obtain Eq.~(\ref{eq2}).

%
%

\section{\label{secB}Bogoliubov transformation}
In this Appendix, we show a way to construct the Bogoliubov transformation to get the energy eigenvalues in bosonic BdG systems. In this case, we must diagonalize the real-space Hamiltonian (\ref{eq1}) by the basis transformation $\left({\bm a}^\dag~{\bm a}\right)=\left({\bm\alpha}^\dag~{\bm\alpha}\right)T^\dag$, where $T$ is the paraunitary matrix defined in Eq.~(\ref{eq5}). Then Eq.~(\ref{eq1}) is diagonalized as
\begin{eqnarray}
H=\frac{1}{2}\left({\bm\alpha}^\dag~{\bm\alpha}\right)\left( \begin{array}{cc}
\Lambda & O       \\
O       & \Lambda
\end{array}\right)\left( \begin{array}{c}
{\bm\alpha} \\
{\bm\alpha}^\dag
\end{array}\right),
\label{eqappB1}
\end{eqnarray}
where
\begin{eqnarray}
T^{-1}\left(\tau_zH_{\rm BdG}\right)T=\left( \begin{array}{cc}
\Lambda & O        \\
O       & -\Lambda
\end{array}\right),
\label{eqappB2}
\end{eqnarray}
and $\Lambda$ is a diagonal matrix, and $\tau_z$ is given in Eq.~(\ref{eq3}). We note that the columns of $T$ are the right eigenvectors of $\tau_zH_{\rm BdG}$. On the other hand, since $\tau_zH_{\rm BdG}$ has the pseudo-Hermiticity~\cite{Kawabata2019} defined as
\begin{equation}
\tau_z\left(\tau_zH_{\rm BdG}\right)^\dag\tau_z=\tau_zH_{\rm BdG},
\label{eqappB3}
\end{equation}
the rows of $T^\dag\tau_z$ are the left eigenvectors of $\tau_zH_{\rm BdG}$. Hence, if a set of eigenvectors of $\tau_zH_{\rm BdG}$ forms the biorthogonal basis~\cite{Brody2014}, we can get the paraunitary matrix $T$. In other words, the constitution of the biorthogonal basis is equivalent to the condition (\ref{eq5}).


\providecommand{\noopsort}[1]{}\providecommand{\singleletter}[1]{#1}%

\begin{thebibliography}{52}%
\makeatletter
\providecommand \@ifxundefined [1]{%
 \@ifx{#1\undefined}
}%
\providecommand \@ifnum [1]{%
 \ifnum #1\expandafter \@firstoftwo
 \else \expandafter \@secondoftwo
 \fi
}%
\providecommand \@ifx [1]{%
 \ifx #1\expandafter \@firstoftwo
 \else \expandafter \@secondoftwo
 \fi
}%
\providecommand \natexlab [1]{#1}%
\providecommand \enquote  [1]{``#1''}%
\providecommand \bibnamefont  [1]{#1}%
\providecommand \bibfnamefont [1]{#1}%
\providecommand \citenamefont [1]{#1}%
\providecommand \href@noop [0]{\@secondoftwo}%
\providecommand \href [0]{\begingroup \@sanitize@url \@href}%
\providecommand \@href[1]{\@@startlink{#1}\@@href}%
\providecommand \@@href[1]{\endgroup#1\@@endlink}%
\providecommand \@sanitize@url [0]{\catcode `\\12\catcode `\$12\catcode
  `\&12\catcode `\#12\catcode `\^12\catcode `\_12\catcode `\%12\relax}%
\providecommand \@@startlink[1]{}%
\providecommand \@@endlink[0]{}%
\providecommand \url  [0]{\begingroup\@sanitize@url \@url }%
\providecommand \@url [1]{\endgroup\@href {#1}{\urlprefix }}%
\providecommand \urlprefix  [0]{URL }%
\providecommand \Eprint [0]{\href }%
\providecommand \doibase [0]{https://doi.org/}%
\providecommand \selectlanguage [0]{\@gobble}%
\providecommand \bibinfo  [0]{\@secondoftwo}%
\providecommand \bibfield  [0]{\@secondoftwo}%
\providecommand \translation [1]{[#1]}%
\providecommand \BibitemOpen [0]{}%
\providecommand \bibitemStop [0]{}%
\providecommand \bibitemNoStop [0]{.\EOS\space}%
\providecommand \EOS [0]{\spacefactor3000\relax}%
\providecommand \BibitemShut  [1]{\csname bibitem#1\endcsname}%
\let\auto@bib@innerbib\@empty
\bibitem [{\citenamefont {Zhang}\ \emph {et~al.}(2013)\citenamefont {Zhang},
  \citenamefont {Ren}, \citenamefont {Wang},\ and\ \citenamefont
  {Li}}]{Zhang2013}%
  \BibitemOpen
  \bibfield  {author} {\bibinfo {author} {\bibfnamefont {L.}~\bibnamefont
  {Zhang}}, \bibinfo {author} {\bibfnamefont {J.}~\bibnamefont {Ren}}, \bibinfo
  {author} {\bibfnamefont {J.-S.}\ \bibnamefont {Wang}},\ and\ \bibinfo
  {author} {\bibfnamefont {B.}~\bibnamefont {Li}},\ }\href
  {https://doi.org/10.1103/PhysRevB.87.144101} {\bibfield  {journal} {\bibinfo
  {journal} {Phys. Rev. B}\ }\textbf {\bibinfo {volume} {87}},\ \bibinfo
  {pages} {144101} (\bibinfo {year} {2013})}\BibitemShut {NoStop}%
\bibitem [{\citenamefont {Shindou}\ \emph
  {et~al.}(2013{\natexlab{a}})\citenamefont {Shindou}, \citenamefont {Ohe},
  \citenamefont {Matsumoto}, \citenamefont {Murakami},\ and\ \citenamefont
  {Saitoh}}]{Shindou2013}%
  \BibitemOpen
  \bibfield  {author} {\bibinfo {author} {\bibfnamefont {R.}~\bibnamefont
  {Shindou}}, \bibinfo {author} {\bibfnamefont {J.-i.}\ \bibnamefont {Ohe}},
  \bibinfo {author} {\bibfnamefont {R.}~\bibnamefont {Matsumoto}}, \bibinfo
  {author} {\bibfnamefont {S.}~\bibnamefont {Murakami}},\ and\ \bibinfo
  {author} {\bibfnamefont {E.}~\bibnamefont {Saitoh}},\ }\href
  {https://doi.org/10.1103/PhysRevB.87.174402} {\bibfield  {journal} {\bibinfo
  {journal} {Phys. Rev. B}\ }\textbf {\bibinfo {volume} {87}},\ \bibinfo
  {pages} {174402} (\bibinfo {year} {2013}{\natexlab{a}})}\BibitemShut
  {NoStop}%
\bibitem [{\citenamefont {Shindou}\ \emph
  {et~al.}(2013{\natexlab{b}})\citenamefont {Shindou}, \citenamefont
  {Matsumoto}, \citenamefont {Murakami},\ and\ \citenamefont
  {Ohe}}]{Shindou2013v2}%
  \BibitemOpen
  \bibfield  {author} {\bibinfo {author} {\bibfnamefont {R.}~\bibnamefont
  {Shindou}}, \bibinfo {author} {\bibfnamefont {R.}~\bibnamefont {Matsumoto}},
  \bibinfo {author} {\bibfnamefont {S.}~\bibnamefont {Murakami}},\ and\
  \bibinfo {author} {\bibfnamefont {J.-i.}\ \bibnamefont {Ohe}},\ }\href
  {https://doi.org/10.1103/PhysRevB.87.174427} {\bibfield  {journal} {\bibinfo
  {journal} {Phys. Rev. B}\ }\textbf {\bibinfo {volume} {87}},\ \bibinfo
  {pages} {174427} (\bibinfo {year} {2013}{\natexlab{b}})}\BibitemShut
  {NoStop}%
\bibitem [{\citenamefont {Matsumoto}\ \emph {et~al.}(2014)\citenamefont
  {Matsumoto}, \citenamefont {Shindou},\ and\ \citenamefont
  {Murakami}}]{Matsumoto2014}%
  \BibitemOpen
  \bibfield  {author} {\bibinfo {author} {\bibfnamefont {R.}~\bibnamefont
  {Matsumoto}}, \bibinfo {author} {\bibfnamefont {R.}~\bibnamefont {Shindou}},\
  and\ \bibinfo {author} {\bibfnamefont {S.}~\bibnamefont {Murakami}},\ }\href
  {https://doi.org/10.1103/PhysRevB.89.054420} {\bibfield  {journal} {\bibinfo
  {journal} {Phys. Rev. B}\ }\textbf {\bibinfo {volume} {89}},\ \bibinfo
  {pages} {054420} (\bibinfo {year} {2014})}\BibitemShut {NoStop}%
\bibitem [{\citenamefont {Zyuzin}\ and\ \citenamefont
  {Kovalev}(2016)}]{Zyuzin2016}%
  \BibitemOpen
  \bibfield  {author} {\bibinfo {author} {\bibfnamefont {V.~A.}\ \bibnamefont
  {Zyuzin}}\ and\ \bibinfo {author} {\bibfnamefont {A.~A.}\ \bibnamefont
  {Kovalev}},\ }\href {https://doi.org/10.1103/PhysRevLett.117.217203}
  {\bibfield  {journal} {\bibinfo  {journal} {Phys. Rev. Lett.}\ }\textbf
  {\bibinfo {volume} {117}},\ \bibinfo {pages} {217203} (\bibinfo {year}
  {2016})}\BibitemShut {NoStop}%
\bibitem [{\citenamefont {Lu}\ and\ \citenamefont {Lu}()}]{Lu2018}%
  \BibitemOpen
  \bibfield  {author} {\bibinfo {author} {\bibfnamefont {F.}~\bibnamefont
  {Lu}}\ and\ \bibinfo {author} {\bibfnamefont {Y.-M.}\ \bibnamefont {Lu}},\
  }\href@noop {} {\bibinfo  {journal} {arXiv:1807.05232}\ }\BibitemShut
  {NoStop}%
\bibitem [{\citenamefont {Kondo}\ \emph
  {et~al.}(2019{\natexlab{a}})\citenamefont {Kondo}, \citenamefont {Akagi},\
  and\ \citenamefont {Katsura}}]{Kondo2019}%
  \BibitemOpen
\bibfield  {journal} {  }\bibfield  {author} {\bibinfo {author} {\bibfnamefont
  {H.}~\bibnamefont {Kondo}}, \bibinfo {author} {\bibfnamefont
  {Y.}~\bibnamefont {Akagi}},\ and\ \bibinfo {author} {\bibfnamefont
  {H.}~\bibnamefont {Katsura}},\ }\href
  {https://doi.org/10.1103/PhysRevB.99.041110} {\bibfield  {journal} {\bibinfo
  {journal} {Phys. Rev. B}\ }\textbf {\bibinfo {volume} {99}},\ \bibinfo
  {pages} {041110(R)} (\bibinfo {year} {2019}{\natexlab{a}})}\BibitemShut
  {NoStop}%
\bibitem [{\citenamefont {Joshi}\ and\ \citenamefont
  {Schnyder}(2019)}]{Joshi2019}%
  \BibitemOpen
  \bibfield  {author} {\bibinfo {author} {\bibfnamefont {D.~G.}\ \bibnamefont
  {Joshi}}\ and\ \bibinfo {author} {\bibfnamefont {A.~P.}\ \bibnamefont
  {Schnyder}},\ }\href {https://doi.org/10.1103/PhysRevB.100.020407} {\bibfield
   {journal} {\bibinfo  {journal} {Phys. Rev. B}\ }\textbf {\bibinfo {volume}
  {100}},\ \bibinfo {pages} {020407(R)} (\bibinfo {year} {2019})}\BibitemShut
  {NoStop}%
\bibitem [{\citenamefont {Kondo}\ \emph
  {et~al.}(2019{\natexlab{b}})\citenamefont {Kondo}, \citenamefont {Akagi},\
  and\ \citenamefont {Katsura}}]{Kondo2019v2}%
  \BibitemOpen
  \bibfield  {author} {\bibinfo {author} {\bibfnamefont {H.}~\bibnamefont
  {Kondo}}, \bibinfo {author} {\bibfnamefont {Y.}~\bibnamefont {Akagi}},\ and\
  \bibinfo {author} {\bibfnamefont {H.}~\bibnamefont {Katsura}},\ }\href
  {https://doi.org/10.1103/PhysRevB.100.144401} {\bibfield  {journal} {\bibinfo
   {journal} {Phys. Rev. B}\ }\textbf {\bibinfo {volume} {100}},\ \bibinfo
  {pages} {144401} (\bibinfo {year} {2019}{\natexlab{b}})}\BibitemShut
  {NoStop}%
\bibitem [{\citenamefont {Hwang}\ \emph {et~al.}(2020)\citenamefont {Hwang},
  \citenamefont {Trivedi},\ and\ \citenamefont {Randeria}}]{Hwang2020}%
  \BibitemOpen
  \bibfield  {author} {\bibinfo {author} {\bibfnamefont {K.}~\bibnamefont
  {Hwang}}, \bibinfo {author} {\bibfnamefont {N.}~\bibnamefont {Trivedi}},\
  and\ \bibinfo {author} {\bibfnamefont {M.}~\bibnamefont {Randeria}},\ }\href
  {https://doi.org/10.1103/PhysRevLett.125.047203} {\bibfield  {journal}
  {\bibinfo  {journal} {Phys. Rev. Lett.}\ }\textbf {\bibinfo {volume} {125}},\
  \bibinfo {pages} {047203} (\bibinfo {year} {2020})}\BibitemShut {NoStop}%
\bibitem [{\citenamefont {Fallani}\ \emph {et~al.}(2004)\citenamefont
  {Fallani}, \citenamefont {De~Sarlo}, \citenamefont {Lye}, \citenamefont
  {Modugno}, \citenamefont {Saers}, \citenamefont {Fort},\ and\ \citenamefont
  {Inguscio}}]{Fallani2004}%
  \BibitemOpen
  \bibfield  {author} {\bibinfo {author} {\bibfnamefont {L.}~\bibnamefont
  {Fallani}}, \bibinfo {author} {\bibfnamefont {L.}~\bibnamefont {De~Sarlo}},
  \bibinfo {author} {\bibfnamefont {J.~E.}\ \bibnamefont {Lye}}, \bibinfo
  {author} {\bibfnamefont {M.}~\bibnamefont {Modugno}}, \bibinfo {author}
  {\bibfnamefont {R.}~\bibnamefont {Saers}}, \bibinfo {author} {\bibfnamefont
  {C.}~\bibnamefont {Fort}},\ and\ \bibinfo {author} {\bibfnamefont
  {M.}~\bibnamefont {Inguscio}},\ }\href
  {https://doi.org/10.1103/PhysRevLett.93.140406} {\bibfield  {journal}
  {\bibinfo  {journal} {Phys. Rev. Lett.}\ }\textbf {\bibinfo {volume} {93}},\
  \bibinfo {pages} {140406} (\bibinfo {year} {2004})}\BibitemShut {NoStop}%
\bibitem [{\citenamefont {Nakamura}\ \emph {et~al.}(2008)\citenamefont
  {Nakamura}, \citenamefont {Mine}, \citenamefont {Okumura},\ and\
  \citenamefont {Yamanaka}}]{Nakamura2008}%
  \BibitemOpen
  \bibfield  {author} {\bibinfo {author} {\bibfnamefont {Y.}~\bibnamefont
  {Nakamura}}, \bibinfo {author} {\bibfnamefont {M.}~\bibnamefont {Mine}},
  \bibinfo {author} {\bibfnamefont {M.}~\bibnamefont {Okumura}},\ and\ \bibinfo
  {author} {\bibfnamefont {Y.}~\bibnamefont {Yamanaka}},\ }\href
  {https://doi.org/10.1103/PhysRevA.77.043601} {\bibfield  {journal} {\bibinfo
  {journal} {Phys. Rev. A}\ }\textbf {\bibinfo {volume} {77}},\ \bibinfo
  {pages} {043601} (\bibinfo {year} {2008})}\BibitemShut {NoStop}%
\bibitem [{\citenamefont {Kawabata}\ \emph {et~al.}(2019)\citenamefont
  {Kawabata}, \citenamefont {Shiozaki}, \citenamefont {Ueda},\ and\
  \citenamefont {Sato}}]{Kawabata2019}%
  \BibitemOpen
  \bibfield  {author} {\bibinfo {author} {\bibfnamefont {K.}~\bibnamefont
  {Kawabata}}, \bibinfo {author} {\bibfnamefont {K.}~\bibnamefont {Shiozaki}},
  \bibinfo {author} {\bibfnamefont {M.}~\bibnamefont {Ueda}},\ and\ \bibinfo
  {author} {\bibfnamefont {M.}~\bibnamefont {Sato}},\ }\href
  {https://doi.org/10.1103/PhysRevX.9.041015} {\bibfield  {journal} {\bibinfo
  {journal} {Phys. Rev. X}\ }\textbf {\bibinfo {volume} {9}},\ \bibinfo {pages}
  {041015} (\bibinfo {year} {2019})}\BibitemShut {NoStop}%
\bibitem [{\citenamefont {Barnett}(2013)}]{Barnett2013}%
  \BibitemOpen
  \bibfield  {author} {\bibinfo {author} {\bibfnamefont {R.}~\bibnamefont
  {Barnett}},\ }\href {https://doi.org/10.1103/PhysRevA.88.063631} {\bibfield
  {journal} {\bibinfo  {journal} {Phys. Rev. A}\ }\textbf {\bibinfo {volume}
  {88}},\ \bibinfo {pages} {063631} (\bibinfo {year} {2013})}\BibitemShut
  {NoStop}%
\bibitem [{\citenamefont {Engelhardt}\ and\ \citenamefont
  {Brandes}(2015)}]{Engelhardt2015}%
  \BibitemOpen
  \bibfield  {author} {\bibinfo {author} {\bibfnamefont {G.}~\bibnamefont
  {Engelhardt}}\ and\ \bibinfo {author} {\bibfnamefont {T.}~\bibnamefont
  {Brandes}},\ }\href {https://doi.org/10.1103/PhysRevA.91.053621} {\bibfield
  {journal} {\bibinfo  {journal} {Phys. Rev. A}\ }\textbf {\bibinfo {volume}
  {91}},\ \bibinfo {pages} {053621} (\bibinfo {year} {2015})}\BibitemShut
  {NoStop}%
\bibitem [{\citenamefont {Furukawa}\ and\ \citenamefont
  {Ueda}(2015)}]{Furukawa2015}%
  \BibitemOpen
  \bibfield  {author} {\bibinfo {author} {\bibfnamefont {S.}~\bibnamefont
  {Furukawa}}\ and\ \bibinfo {author} {\bibfnamefont {M.}~\bibnamefont
  {Ueda}},\ }\href@noop {} {\bibfield  {journal} {\bibinfo  {journal} {New J.
  Phys.}\ }\textbf {\bibinfo {volume} {17}},\ \bibinfo {pages} {115014}
  (\bibinfo {year} {2015})}\BibitemShut {NoStop}%
\bibitem [{\citenamefont {Galilo}\ \emph {et~al.}(2015)\citenamefont {Galilo},
  \citenamefont {Lee},\ and\ \citenamefont {Barnett}}]{Galilo2015}%
  \BibitemOpen
  \bibfield  {author} {\bibinfo {author} {\bibfnamefont {B.}~\bibnamefont
  {Galilo}}, \bibinfo {author} {\bibfnamefont {D.~K.~K.}\ \bibnamefont {Lee}},\
  and\ \bibinfo {author} {\bibfnamefont {R.}~\bibnamefont {Barnett}},\ }\href
  {https://doi.org/10.1103/PhysRevLett.115.245302} {\bibfield  {journal}
  {\bibinfo  {journal} {Phys. Rev. Lett.}\ }\textbf {\bibinfo {volume} {115}},\
  \bibinfo {pages} {245302} (\bibinfo {year} {2015})}\BibitemShut {NoStop}%
\bibitem [{\citenamefont {Engelhardt}\ \emph {et~al.}(2016)\citenamefont
  {Engelhardt}, \citenamefont {Benito}, \citenamefont {Platero},\ and\
  \citenamefont {Brandes}}]{Engelhardt2016}%
  \BibitemOpen
  \bibfield  {author} {\bibinfo {author} {\bibfnamefont {G.}~\bibnamefont
  {Engelhardt}}, \bibinfo {author} {\bibfnamefont {M.}~\bibnamefont {Benito}},
  \bibinfo {author} {\bibfnamefont {G.}~\bibnamefont {Platero}},\ and\ \bibinfo
  {author} {\bibfnamefont {T.}~\bibnamefont {Brandes}},\ }\href
  {https://doi.org/10.1103/PhysRevLett.117.045302} {\bibfield  {journal}
  {\bibinfo  {journal} {Phys. Rev. Lett.}\ }\textbf {\bibinfo {volume} {117}},\
  \bibinfo {pages} {045302} (\bibinfo {year} {2016})}\BibitemShut {NoStop}%
\bibitem [{\citenamefont {McDonald}\ \emph {et~al.}(2018)\citenamefont
  {McDonald}, \citenamefont {Pereg-Barnea},\ and\ \citenamefont
  {Clerk}}]{McDonald2018}%
  \BibitemOpen
  \bibfield  {author} {\bibinfo {author} {\bibfnamefont {A.}~\bibnamefont
  {McDonald}}, \bibinfo {author} {\bibfnamefont {T.}~\bibnamefont
  {Pereg-Barnea}},\ and\ \bibinfo {author} {\bibfnamefont {A.~A.}\ \bibnamefont
  {Clerk}},\ }\href {https://doi.org/10.1103/PhysRevX.8.041031} {\bibfield
  {journal} {\bibinfo  {journal} {Phys. Rev. X}\ }\textbf {\bibinfo {volume}
  {8}},\ \bibinfo {pages} {041031} (\bibinfo {year} {2018})}\BibitemShut
  {NoStop}%
\bibitem [{\citenamefont {Yao}\ and\ \citenamefont {Wang}(2018)}]{Yao2018}%
  \BibitemOpen
  \bibfield  {author} {\bibinfo {author} {\bibfnamefont {S.}~\bibnamefont
  {Yao}}\ and\ \bibinfo {author} {\bibfnamefont {Z.}~\bibnamefont {Wang}},\
  }\href {https://doi.org/10.1103/PhysRevLett.121.086803} {\bibfield  {journal}
  {\bibinfo  {journal} {Phys. Rev. Lett.}\ }\textbf {\bibinfo {volume} {121}},\
  \bibinfo {pages} {086803} (\bibinfo {year} {2018})}\BibitemShut {NoStop}%
\bibitem [{\citenamefont {Jin}\ and\ \citenamefont {Song}(2019)}]{Jin2019}%
  \BibitemOpen
  \bibfield  {author} {\bibinfo {author} {\bibfnamefont {L.}~\bibnamefont
  {Jin}}\ and\ \bibinfo {author} {\bibfnamefont {Z.}~\bibnamefont {Song}},\
  }\href {https://doi.org/10.1103/PhysRevB.99.081103} {\bibfield  {journal}
  {\bibinfo  {journal} {Phys. Rev. B}\ }\textbf {\bibinfo {volume} {99}},\
  \bibinfo {pages} {081103(R)} (\bibinfo {year} {2019})}\BibitemShut {NoStop}%
\bibitem [{\citenamefont {Deng}\ and\ \citenamefont {Yi}(2019)}]{Deng2019}%
  \BibitemOpen
  \bibfield  {author} {\bibinfo {author} {\bibfnamefont {T.-S.}\ \bibnamefont
  {Deng}}\ and\ \bibinfo {author} {\bibfnamefont {W.}~\bibnamefont {Yi}},\
  }\href {https://doi.org/10.1103/PhysRevB.100.035102} {\bibfield  {journal}
  {\bibinfo  {journal} {Phys. Rev. B}\ }\textbf {\bibinfo {volume} {100}},\
  \bibinfo {pages} {035102} (\bibinfo {year} {2019})}\BibitemShut {NoStop}%
\bibitem [{\citenamefont {Okuma}\ \emph {et~al.}(2020)\citenamefont {Okuma},
  \citenamefont {Kawabata}, \citenamefont {Shiozaki},\ and\ \citenamefont
  {Sato}}]{Okuma2020}%
  \BibitemOpen
  \bibfield  {author} {\bibinfo {author} {\bibfnamefont {N.}~\bibnamefont
  {Okuma}}, \bibinfo {author} {\bibfnamefont {K.}~\bibnamefont {Kawabata}},
  \bibinfo {author} {\bibfnamefont {K.}~\bibnamefont {Shiozaki}},\ and\
  \bibinfo {author} {\bibfnamefont {M.}~\bibnamefont {Sato}},\ }\href
  {https://doi.org/10.1103/PhysRevLett.124.086801} {\bibfield  {journal}
  {\bibinfo  {journal} {Phys. Rev. Lett.}\ }\textbf {\bibinfo {volume} {124}},\
  \bibinfo {pages} {086801} (\bibinfo {year} {2020})}\BibitemShut {NoStop}%
\bibitem [{\citenamefont {Borgnia}\ \emph {et~al.}(2020)\citenamefont
  {Borgnia}, \citenamefont {Kruchkov},\ and\ \citenamefont
  {Slager}}]{Borgnia2020}%
  \BibitemOpen
  \bibfield  {author} {\bibinfo {author} {\bibfnamefont {D.~S.}\ \bibnamefont
  {Borgnia}}, \bibinfo {author} {\bibfnamefont {A.~J.}\ \bibnamefont
  {Kruchkov}},\ and\ \bibinfo {author} {\bibfnamefont {R.-J.}\ \bibnamefont
  {Slager}},\ }\href {https://doi.org/10.1103/PhysRevLett.124.056802}
  {\bibfield  {journal} {\bibinfo  {journal} {Phys. Rev. Lett.}\ }\textbf
  {\bibinfo {volume} {124}},\ \bibinfo {pages} {056802} (\bibinfo {year}
  {2020})}\BibitemShut {NoStop}%
\bibitem [{\citenamefont {Yu}\ \emph {et~al.}(2020{\natexlab{a}})\citenamefont
  {Yu}, \citenamefont {Zhang}, \citenamefont {Sharma}, \citenamefont {Zhang},
  \citenamefont {Blanter},\ and\ \citenamefont {Bauer}}]{Yu2020}%
  \BibitemOpen
  \bibfield  {author} {\bibinfo {author} {\bibfnamefont {T.}~\bibnamefont
  {Yu}}, \bibinfo {author} {\bibfnamefont {Y.-X.}\ \bibnamefont {Zhang}},
  \bibinfo {author} {\bibfnamefont {S.}~\bibnamefont {Sharma}}, \bibinfo
  {author} {\bibfnamefont {X.}~\bibnamefont {Zhang}}, \bibinfo {author}
  {\bibfnamefont {Y.~M.}\ \bibnamefont {Blanter}},\ and\ \bibinfo {author}
  {\bibfnamefont {G.~E.~W.}\ \bibnamefont {Bauer}},\ }\href
  {https://doi.org/10.1103/PhysRevLett.124.107202} {\bibfield  {journal}
  {\bibinfo  {journal} {Phys. Rev. Lett.}\ }\textbf {\bibinfo {volume} {124}},\
  \bibinfo {pages} {107202} (\bibinfo {year} {2020}{\natexlab{a}})}\BibitemShut
  {NoStop}%
\bibitem [{\citenamefont {Yu}\ \emph {et~al.}(2020{\natexlab{b}})\citenamefont
  {Yu}, \citenamefont {Zhang}, \citenamefont {Sharma}, \citenamefont
  {Blanter},\ and\ \citenamefont {Bauer}}]{Yu2020v2}%
  \BibitemOpen
  \bibfield  {author} {\bibinfo {author} {\bibfnamefont {T.}~\bibnamefont
  {Yu}}, \bibinfo {author} {\bibfnamefont {X.}~\bibnamefont {Zhang}}, \bibinfo
  {author} {\bibfnamefont {S.}~\bibnamefont {Sharma}}, \bibinfo {author}
  {\bibfnamefont {Y.~M.}\ \bibnamefont {Blanter}},\ and\ \bibinfo {author}
  {\bibfnamefont {G.~E.~W.}\ \bibnamefont {Bauer}},\ }\href
  {https://doi.org/10.1103/PhysRevB.101.094414} {\bibfield  {journal} {\bibinfo
   {journal} {Phys. Rev. B}\ }\textbf {\bibinfo {volume} {101}},\ \bibinfo
  {pages} {094414} (\bibinfo {year} {2020}{\natexlab{b}})}\BibitemShut
  {NoStop}%
\bibitem [{\citenamefont {Lee}\ \emph {et~al.}(2020)\citenamefont {Lee},
  \citenamefont {Lee},\ and\ \citenamefont {Yang}}]{Lee2020}%
  \BibitemOpen
  \bibfield  {author} {\bibinfo {author} {\bibfnamefont {E.}~\bibnamefont
  {Lee}}, \bibinfo {author} {\bibfnamefont {H.}~\bibnamefont {Lee}},\ and\
  \bibinfo {author} {\bibfnamefont {B.-J.}\ \bibnamefont {Yang}},\ }\href
  {https://doi.org/10.1103/PhysRevB.101.121109} {\bibfield  {journal} {\bibinfo
   {journal} {Phys. Rev. B}\ }\textbf {\bibinfo {volume} {101}},\ \bibinfo
  {pages} {121109(R)} (\bibinfo {year} {2020})}\BibitemShut {NoStop}%
\bibitem [{\citenamefont {Kawabata}\ \emph
  {et~al.}(2020{\natexlab{a}})\citenamefont {Kawabata}, \citenamefont {Okuma},\
  and\ \citenamefont {Sato}}]{Kawabata2020}%
  \BibitemOpen
  \bibfield  {author} {\bibinfo {author} {\bibfnamefont {K.}~\bibnamefont
  {Kawabata}}, \bibinfo {author} {\bibfnamefont {N.}~\bibnamefont {Okuma}},\
  and\ \bibinfo {author} {\bibfnamefont {M.}~\bibnamefont {Sato}},\ }\href
  {https://doi.org/10.1103/PhysRevB.101.195147} {\bibfield  {journal} {\bibinfo
   {journal} {Phys. Rev. B}\ }\textbf {\bibinfo {volume} {101}},\ \bibinfo
  {pages} {195147} (\bibinfo {year} {2020}{\natexlab{a}})}\BibitemShut
  {NoStop}%
\bibitem [{\citenamefont {Yoshida}\ \emph {et~al.}(2020)\citenamefont
  {Yoshida}, \citenamefont {Mizoguchi},\ and\ \citenamefont
  {Hatsugai}}]{Yoshida2020}%
  \BibitemOpen
  \bibfield  {author} {\bibinfo {author} {\bibfnamefont {T.}~\bibnamefont
  {Yoshida}}, \bibinfo {author} {\bibfnamefont {T.}~\bibnamefont {Mizoguchi}},\
  and\ \bibinfo {author} {\bibfnamefont {Y.}~\bibnamefont {Hatsugai}},\ }\href
  {https://doi.org/10.1103/PhysRevResearch.2.022062} {\bibfield  {journal}
  {\bibinfo  {journal} {Phys. Rev. Research}\ }\textbf {\bibinfo {volume}
  {2}},\ \bibinfo {pages} {022062(R)} (\bibinfo {year} {2020})}\BibitemShut
  {NoStop}%
\bibitem [{\citenamefont {Zhang}\ \emph {et~al.}(2020)\citenamefont {Zhang},
  \citenamefont {Yang},\ and\ \citenamefont {Fang}}]{Zhang2020}%
  \BibitemOpen
  \bibfield  {author} {\bibinfo {author} {\bibfnamefont {K.}~\bibnamefont
  {Zhang}}, \bibinfo {author} {\bibfnamefont {Z.}~\bibnamefont {Yang}},\ and\
  \bibinfo {author} {\bibfnamefont {C.}~\bibnamefont {Fang}},\ }\href
  {https://doi.org/10.1103/PhysRevLett.125.126402} {\bibfield  {journal}
  {\bibinfo  {journal} {Phys. Rev. Lett.}\ }\textbf {\bibinfo {volume} {125}},\
  \bibinfo {pages} {126402} (\bibinfo {year} {2020})}\BibitemShut {NoStop}%
\bibitem [{\citenamefont {Yi}\ and\ \citenamefont {Yang}(2020)}]{Yi2020}%
  \BibitemOpen
  \bibfield  {author} {\bibinfo {author} {\bibfnamefont {Y.}~\bibnamefont
  {Yi}}\ and\ \bibinfo {author} {\bibfnamefont {Z.}~\bibnamefont {Yang}},\
  }\href {https://doi.org/10.1103/PhysRevLett.125.186802} {\bibfield  {journal}
  {\bibinfo  {journal} {Phys. Rev. Lett.}\ }\textbf {\bibinfo {volume} {125}},\
  \bibinfo {pages} {186802} (\bibinfo {year} {2020})}\BibitemShut {NoStop}%
\bibitem [{\citenamefont {Li}\ \emph {et~al.}(2020)\citenamefont {Li},
  \citenamefont {Lee}, \citenamefont {Mu},\ and\ \citenamefont
  {Gong}}]{Li2020}%
  \BibitemOpen
  \bibfield  {author} {\bibinfo {author} {\bibfnamefont {L.}~\bibnamefont
  {Li}}, \bibinfo {author} {\bibfnamefont {C.~H.}\ \bibnamefont {Lee}},
  \bibinfo {author} {\bibfnamefont {S.}~\bibnamefont {Mu}},\ and\ \bibinfo
  {author} {\bibfnamefont {J.}~\bibnamefont {Gong}},\ }\href@noop {} {\bibfield
   {journal} {\bibinfo  {journal} {Nat. Commun.}\ }\textbf {\bibinfo {volume}
  {11}},\ \bibinfo {pages} {5491} (\bibinfo {year} {2020})}\BibitemShut
  {NoStop}%
\bibitem [{\citenamefont {Kawabata}\ \emph
  {et~al.}(2020{\natexlab{b}})\citenamefont {Kawabata}, \citenamefont {Sato},\
  and\ \citenamefont {Shiozaki}}]{Kawabata2020v2}%
  \BibitemOpen
  \bibfield  {author} {\bibinfo {author} {\bibfnamefont {K.}~\bibnamefont
  {Kawabata}}, \bibinfo {author} {\bibfnamefont {M.}~\bibnamefont {Sato}},\
  and\ \bibinfo {author} {\bibfnamefont {K.}~\bibnamefont {Shiozaki}},\ }\href
  {https://doi.org/10.1103/PhysRevB.102.205118} {\bibfield  {journal} {\bibinfo
   {journal} {Phys. Rev. B}\ }\textbf {\bibinfo {volume} {102}},\ \bibinfo
  {pages} {205118} (\bibinfo {year} {2020}{\natexlab{b}})}\BibitemShut
  {NoStop}%
\bibitem [{\citenamefont {Okugawa}\ \emph {et~al.}(2020)\citenamefont
  {Okugawa}, \citenamefont {Takahashi},\ and\ \citenamefont
  {Yokomizo}}]{Okugawa2020}%
  \BibitemOpen
  \bibfield  {author} {\bibinfo {author} {\bibfnamefont {R.}~\bibnamefont
  {Okugawa}}, \bibinfo {author} {\bibfnamefont {R.}~\bibnamefont {Takahashi}},\
  and\ \bibinfo {author} {\bibfnamefont {K.}~\bibnamefont {Yokomizo}},\ }\href
  {https://doi.org/10.1103/PhysRevB.102.241202} {\bibfield  {journal} {\bibinfo
   {journal} {Phys. Rev. B}\ }\textbf {\bibinfo {volume} {102}},\ \bibinfo
  {pages} {241202(R)} (\bibinfo {year} {2020})}\BibitemShut {NoStop}%
\bibitem [{\citenamefont {Brandenbourger}\ \emph {et~al.}(2019)\citenamefont
  {Brandenbourger}, \citenamefont {Locsin}, \citenamefont {Lerner},\ and\
  \citenamefont {Coulais}}]{Brandenbourger2019}%
  \BibitemOpen
  \bibfield  {author} {\bibinfo {author} {\bibfnamefont {M.}~\bibnamefont
  {Brandenbourger}}, \bibinfo {author} {\bibfnamefont {X.}~\bibnamefont
  {Locsin}}, \bibinfo {author} {\bibfnamefont {E.}~\bibnamefont {Lerner}},\
  and\ \bibinfo {author} {\bibfnamefont {C.}~\bibnamefont {Coulais}},\
  }\href@noop {} {\bibfield  {journal} {\bibinfo  {journal} {Nat. Commun.}\
  }\textbf {\bibinfo {volume} {10}},\ \bibinfo {pages} {4608} (\bibinfo {year}
  {2019})}\BibitemShut {NoStop}%
\bibitem [{\citenamefont {Gou}\ \emph {et~al.}(2020)\citenamefont {Gou},
  \citenamefont {Chen}, \citenamefont {Xie}, \citenamefont {Xiao},
  \citenamefont {Deng}, \citenamefont {Gadway}, \citenamefont {Yi},\ and\
  \citenamefont {Yan}}]{Gou2020}%
  \BibitemOpen
  \bibfield  {author} {\bibinfo {author} {\bibfnamefont {W.}~\bibnamefont
  {Gou}}, \bibinfo {author} {\bibfnamefont {T.}~\bibnamefont {Chen}}, \bibinfo
  {author} {\bibfnamefont {D.}~\bibnamefont {Xie}}, \bibinfo {author}
  {\bibfnamefont {T.}~\bibnamefont {Xiao}}, \bibinfo {author} {\bibfnamefont
  {T.-S.}\ \bibnamefont {Deng}}, \bibinfo {author} {\bibfnamefont
  {B.}~\bibnamefont {Gadway}}, \bibinfo {author} {\bibfnamefont
  {W.}~\bibnamefont {Yi}},\ and\ \bibinfo {author} {\bibfnamefont
  {B.}~\bibnamefont {Yan}},\ }\href
  {https://doi.org/10.1103/PhysRevLett.124.070402} {\bibfield  {journal}
  {\bibinfo  {journal} {Phys. Rev. Lett.}\ }\textbf {\bibinfo {volume} {124}},\
  \bibinfo {pages} {070402} (\bibinfo {year} {2020})}\BibitemShut {NoStop}%
\bibitem [{\citenamefont {Xiao}\ \emph {et~al.}(2020)\citenamefont {Xiao},
  \citenamefont {Deng}, \citenamefont {Wang}, \citenamefont {Zhu},
  \citenamefont {Wang}, \citenamefont {Yi},\ and\ \citenamefont
  {Xue}}]{Xiao2020}%
  \BibitemOpen
  \bibfield  {author} {\bibinfo {author} {\bibfnamefont {L.}~\bibnamefont
  {Xiao}}, \bibinfo {author} {\bibfnamefont {T.}~\bibnamefont {Deng}}, \bibinfo
  {author} {\bibfnamefont {K.}~\bibnamefont {Wang}}, \bibinfo {author}
  {\bibfnamefont {G.}~\bibnamefont {Zhu}}, \bibinfo {author} {\bibfnamefont
  {Z.}~\bibnamefont {Wang}}, \bibinfo {author} {\bibfnamefont {W.}~\bibnamefont
  {Yi}},\ and\ \bibinfo {author} {\bibfnamefont {P.}~\bibnamefont {Xue}},\
  }\href@noop {} {\bibfield  {journal} {\bibinfo  {journal} {Nat. Phys.}\
  }\textbf {\bibinfo {volume} {16}},\ \bibinfo {pages} {761} (\bibinfo {year}
  {2020})}\BibitemShut {NoStop}%
\bibitem [{\citenamefont {Hofmann}\ \emph {et~al.}(2020)\citenamefont
  {Hofmann}, \citenamefont {Helbig}, \citenamefont {Schindler}, \citenamefont
  {Salgo}, \citenamefont {Brzezi\ifmmode~\acute{n}\else \'{n}\fi{}ska},
  \citenamefont {Greiter}, \citenamefont {Kiessling}, \citenamefont {Wolf},
  \citenamefont {Vollhardt}, \citenamefont {Kaba\ifmmode~\check{s}\else
  \v{s}\fi{}i}, \citenamefont {Lee}, \citenamefont {Bilu\ifmmode \check{s}\else
  \v{s}\fi{}i\ifmmode~\acute{c}\else \'{c}\fi{}}, \citenamefont {Thomale},\
  and\ \citenamefont {Neupert}}]{Hofmann2020}%
  \BibitemOpen
  \bibfield  {author} {\bibinfo {author} {\bibfnamefont {T.}~\bibnamefont
  {Hofmann}}, \bibinfo {author} {\bibfnamefont {T.}~\bibnamefont {Helbig}},
  \bibinfo {author} {\bibfnamefont {F.}~\bibnamefont {Schindler}}, \bibinfo
  {author} {\bibfnamefont {N.}~\bibnamefont {Salgo}}, \bibinfo {author}
  {\bibfnamefont {M.}~\bibnamefont {Brzezi\ifmmode~\acute{n}\else
  \'{n}\fi{}ska}}, \bibinfo {author} {\bibfnamefont {M.}~\bibnamefont
  {Greiter}}, \bibinfo {author} {\bibfnamefont {T.}~\bibnamefont {Kiessling}},
  \bibinfo {author} {\bibfnamefont {D.}~\bibnamefont {Wolf}}, \bibinfo {author}
  {\bibfnamefont {A.}~\bibnamefont {Vollhardt}}, \bibinfo {author}
  {\bibfnamefont {A.}~\bibnamefont {Kaba\ifmmode~\check{s}\else \v{s}\fi{}i}},
  \bibinfo {author} {\bibfnamefont {C.~H.}\ \bibnamefont {Lee}}, \bibinfo
  {author} {\bibfnamefont {A.}~\bibnamefont {Bilu\ifmmode \check{s}\else
  \v{s}\fi{}i\ifmmode~\acute{c}\else \'{c}\fi{}}}, \bibinfo {author}
  {\bibfnamefont {R.}~\bibnamefont {Thomale}},\ and\ \bibinfo {author}
  {\bibfnamefont {T.}~\bibnamefont {Neupert}},\ }\href
  {https://doi.org/10.1103/PhysRevResearch.2.023265} {\bibfield  {journal}
  {\bibinfo  {journal} {Phys. Rev. Research}\ }\textbf {\bibinfo {volume}
  {2}},\ \bibinfo {pages} {023265} (\bibinfo {year} {2020})}\BibitemShut
  {NoStop}%
\bibitem [{\citenamefont {Helbig}\ \emph {et~al.}(2020)\citenamefont {Helbig},
  \citenamefont {Hofmann}, \citenamefont {Imhof}, \citenamefont {Abdelghany},
  \citenamefont {Kiessling}, \citenamefont {Molenkamp}, \citenamefont {Lee},
  \citenamefont {Szameit}, \citenamefont {Greiter},\ and\ \citenamefont
  {Thomale}}]{Helbig2020}%
  \BibitemOpen
  \bibfield  {author} {\bibinfo {author} {\bibfnamefont {T.}~\bibnamefont
  {Helbig}}, \bibinfo {author} {\bibfnamefont {T.}~\bibnamefont {Hofmann}},
  \bibinfo {author} {\bibfnamefont {S.}~\bibnamefont {Imhof}}, \bibinfo
  {author} {\bibfnamefont {M.}~\bibnamefont {Abdelghany}}, \bibinfo {author}
  {\bibfnamefont {T.}~\bibnamefont {Kiessling}}, \bibinfo {author}
  {\bibfnamefont {L.}~\bibnamefont {Molenkamp}}, \bibinfo {author}
  {\bibfnamefont {C.}~\bibnamefont {Lee}}, \bibinfo {author} {\bibfnamefont
  {A.}~\bibnamefont {Szameit}}, \bibinfo {author} {\bibfnamefont
  {M.}~\bibnamefont {Greiter}},\ and\ \bibinfo {author} {\bibfnamefont
  {R.}~\bibnamefont {Thomale}},\ }\href@noop {} {\bibfield  {journal} {\bibinfo
   {journal} {Nat. Phys.}\ }\textbf {\bibinfo {volume} {16}},\ \bibinfo {pages}
  {747} (\bibinfo {year} {2020})}\BibitemShut {NoStop}%
\bibitem [{\citenamefont {Yokomizo}\ and\ \citenamefont
  {Murakami}(2019)}]{Yokomizo2019}%
  \BibitemOpen
  \bibfield  {author} {\bibinfo {author} {\bibfnamefont {K.}~\bibnamefont
  {Yokomizo}}\ and\ \bibinfo {author} {\bibfnamefont {S.}~\bibnamefont
  {Murakami}},\ }\href {https://doi.org/10.1103/PhysRevLett.123.066404}
  {\bibfield  {journal} {\bibinfo  {journal} {Phys. Rev. Lett.}\ }\textbf
  {\bibinfo {volume} {123}},\ \bibinfo {pages} {066404} (\bibinfo {year}
  {2019})}\BibitemShut {NoStop}%
\bibitem [{\citenamefont {Yokomizo}\ and\ \citenamefont
  {Shuichi}(2020)}]{Yokomizo2020}%
  \BibitemOpen
  \bibfield  {author} {\bibinfo {author} {\bibfnamefont {K.}~\bibnamefont
  {Yokomizo}}\ and\ \bibinfo {author} {\bibfnamefont {M.}~\bibnamefont
  {Shuichi}},\ }\href@noop {} {\bibfield  {journal} {\bibinfo  {journal} {Prog.
  Theor. Exp. Phys.}\ }\textbf {\bibinfo {volume} {2020}},\ \bibinfo {pages}
  {12A102} (\bibinfo {year} {2020})}\BibitemShut {NoStop}%
\bibitem [{\citenamefont {Yokomizo}\ and\ \citenamefont
  {Murakami}(2020)}]{Yokomizo2020v2}%
  \BibitemOpen
  \bibfield  {author} {\bibinfo {author} {\bibfnamefont {K.}~\bibnamefont
  {Yokomizo}}\ and\ \bibinfo {author} {\bibfnamefont {S.}~\bibnamefont
  {Murakami}},\ }\href {https://doi.org/10.1103/PhysRevResearch.2.043045}
  {\bibfield  {journal} {\bibinfo  {journal} {Phys. Rev. Research}\ }\textbf
  {\bibinfo {volume} {2}},\ \bibinfo {pages} {043045} (\bibinfo {year}
  {2020})}\BibitemShut {NoStop}%
\bibitem [{\citenamefont {Colpa}(1978)}]{Colpa1978}%
  \BibitemOpen
  \bibfield  {author} {\bibinfo {author} {\bibfnamefont {J.}~\bibnamefont
  {Colpa}},\ }\href@noop {} {\bibfield  {journal} {\bibinfo  {journal} {Physica
  A}\ }\textbf {\bibinfo {volume} {93}},\ \bibinfo {pages} {327} (\bibinfo
  {year} {1978})}\BibitemShut {NoStop}%
\bibitem [{\citenamefont {Lieu}(2018)}]{Lieu2018}%
  \BibitemOpen
  \bibfield  {author} {\bibinfo {author} {\bibfnamefont {S.}~\bibnamefont
  {Lieu}},\ }\href {https://doi.org/10.1103/PhysRevB.98.115135} {\bibfield
  {journal} {\bibinfo  {journal} {Phys. Rev. B}\ }\textbf {\bibinfo {volume}
  {98}},\ \bibinfo {pages} {115135} (\bibinfo {year} {2018})}\BibitemShut
  {NoStop}%
\bibitem [{\citenamefont {Yang}\ \emph {et~al.}(2020)\citenamefont {Yang},
  \citenamefont {Zhang}, \citenamefont {Fang},\ and\ \citenamefont
  {Hu}}]{Yang2020}%
  \BibitemOpen
  \bibfield  {author} {\bibinfo {author} {\bibfnamefont {Z.}~\bibnamefont
  {Yang}}, \bibinfo {author} {\bibfnamefont {K.}~\bibnamefont {Zhang}},
  \bibinfo {author} {\bibfnamefont {C.}~\bibnamefont {Fang}},\ and\ \bibinfo
  {author} {\bibfnamefont {J.}~\bibnamefont {Hu}},\ }\href
  {https://doi.org/10.1103/PhysRevLett.125.226402} {\bibfield  {journal}
  {\bibinfo  {journal} {Phys. Rev. Lett.}\ }\textbf {\bibinfo {volume} {125}},\
  \bibinfo {pages} {226402} (\bibinfo {year} {2020})}\BibitemShut {NoStop}%
\bibitem [{\citenamefont {Hatano}\ and\ \citenamefont
  {Nelson}(1996)}]{Hatano1996}%
  \BibitemOpen
  \bibfield  {author} {\bibinfo {author} {\bibfnamefont {N.}~\bibnamefont
  {Hatano}}\ and\ \bibinfo {author} {\bibfnamefont {D.~R.}\ \bibnamefont
  {Nelson}},\ }\href {https://doi.org/10.1103/PhysRevLett.77.570} {\bibfield
  {journal} {\bibinfo  {journal} {Phys. Rev. Lett.}\ }\textbf {\bibinfo
  {volume} {77}},\ \bibinfo {pages} {570} (\bibinfo {year} {1996})}\BibitemShut
  {NoStop}%
\bibitem [{\citenamefont {Okuma}\ and\ \citenamefont {Sato}(2019)}]{Okuma2019}%
  \BibitemOpen
  \bibfield  {author} {\bibinfo {author} {\bibfnamefont {N.}~\bibnamefont
  {Okuma}}\ and\ \bibinfo {author} {\bibfnamefont {M.}~\bibnamefont {Sato}},\
  }\href {https://doi.org/10.1103/PhysRevLett.123.097701} {\bibfield  {journal}
  {\bibinfo  {journal} {Phys. Rev. Lett.}\ }\textbf {\bibinfo {volume} {123}},\
  \bibinfo {pages} {097701} (\bibinfo {year} {2019})}\BibitemShut {NoStop}%
\bibitem [{\citenamefont {Anderson}\ \emph {et~al.}(1995)\citenamefont
  {Anderson}, \citenamefont {Ensher}, \citenamefont {Matthews}, \citenamefont
  {Wieman},\ and\ \citenamefont {Cornell}}]{Anderson1995}%
  \BibitemOpen
  \bibfield  {author} {\bibinfo {author} {\bibfnamefont {M.~H.}\ \bibnamefont
  {Anderson}}, \bibinfo {author} {\bibfnamefont {J.~R.}\ \bibnamefont
  {Ensher}}, \bibinfo {author} {\bibfnamefont {M.~R.}\ \bibnamefont
  {Matthews}}, \bibinfo {author} {\bibfnamefont {C.~E.}\ \bibnamefont
  {Wieman}},\ and\ \bibinfo {author} {\bibfnamefont {E.~A.}\ \bibnamefont
  {Cornell}},\ }\href@noop {} {\bibfield  {journal} {\bibinfo  {journal}
  {Science}\ }\textbf {\bibinfo {volume} {269}},\ \bibinfo {pages} {198}
  (\bibinfo {year} {1995})}\BibitemShut {NoStop}%
\bibitem [{\citenamefont {Demokritov}\ \emph {et~al.}(2006)\citenamefont
  {Demokritov}, \citenamefont {Demidov}, \citenamefont {Dzyapko}, \citenamefont
  {Melkov}, \citenamefont {Serga}, \citenamefont {Hillebrands},\ and\
  \citenamefont {Slavin}}]{Demokritov2006}%
  \BibitemOpen
  \bibfield  {author} {\bibinfo {author} {\bibfnamefont {S.~O.}\ \bibnamefont
  {Demokritov}}, \bibinfo {author} {\bibfnamefont {V.~E.}\ \bibnamefont
  {Demidov}}, \bibinfo {author} {\bibfnamefont {O.}~\bibnamefont {Dzyapko}},
  \bibinfo {author} {\bibfnamefont {G.~A.}\ \bibnamefont {Melkov}}, \bibinfo
  {author} {\bibfnamefont {A.~A.}\ \bibnamefont {Serga}}, \bibinfo {author}
  {\bibfnamefont {B.}~\bibnamefont {Hillebrands}},\ and\ \bibinfo {author}
  {\bibfnamefont {A.~N.}\ \bibnamefont {Slavin}},\ }\href@noop {} {\bibfield
  {journal} {\bibinfo  {journal} {Nature}\ }\textbf {\bibinfo {volume} {443}},\
  \bibinfo {pages} {430} (\bibinfo {year} {2006})}\BibitemShut {NoStop}%
\bibitem [{\citenamefont {Kondo}\ \emph {et~al.}(2020)\citenamefont {Kondo},
  \citenamefont {Akagi},\ and\ \citenamefont {Katsura}}]{Kondo2020}%
  \BibitemOpen
  \bibfield  {author} {\bibinfo {author} {\bibfnamefont {H.}~\bibnamefont
  {Kondo}}, \bibinfo {author} {\bibfnamefont {Y.}~\bibnamefont {Akagi}},\ and\
  \bibinfo {author} {\bibfnamefont {H.}~\bibnamefont {Katsura}},\ }\href@noop
  {} {\bibfield  {journal} {\bibinfo  {journal} {Prog. Theor. Exp. Phys.}\
  }\textbf {\bibinfo {volume} {2020}},\ \bibinfo {pages} {12A104} (\bibinfo
  {year} {2020})}\BibitemShut {NoStop}%
\bibitem [{\citenamefont {Wang}\ \emph {et~al.}(2021)\citenamefont {Wang},
  \citenamefont {Dutt}, \citenamefont {Yang}, \citenamefont {Wojcik},
  \citenamefont {Vu{\v{c}}kovi{\'c}},\ and\ \citenamefont {Fan}}]{Wang2021}%
  \BibitemOpen
  \bibfield  {author} {\bibinfo {author} {\bibfnamefont {K.}~\bibnamefont
  {Wang}}, \bibinfo {author} {\bibfnamefont {A.}~\bibnamefont {Dutt}}, \bibinfo
  {author} {\bibfnamefont {K.~Y.}\ \bibnamefont {Yang}}, \bibinfo {author}
  {\bibfnamefont {C.~C.}\ \bibnamefont {Wojcik}}, \bibinfo {author}
  {\bibfnamefont {J.}~\bibnamefont {Vu{\v{c}}kovi{\'c}}},\ and\ \bibinfo
  {author} {\bibfnamefont {S.}~\bibnamefont {Fan}},\ }\href@noop {} {\bibfield
  {journal} {\bibinfo  {journal} {Science}\ }\textbf {\bibinfo {volume}
  {371}},\ \bibinfo {pages} {1240} (\bibinfo {year} {2021})}\BibitemShut
  {NoStop}%
\bibitem [{\citenamefont {Brody}(2014)}]{Brody2014}%
  \BibitemOpen
  \bibfield  {author} {\bibinfo {author} {\bibfnamefont {D.~C.}\ \bibnamefont
  {Brody}},\ }\href@noop {} {\bibfield  {journal} {\bibinfo  {journal} {J.
  Phys. A: Math. Theor.}\ }\textbf {\bibinfo {volume} {47}},\ \bibinfo {pages}
  {035305} (\bibinfo {year} {2014})}\BibitemShut {NoStop}%
\end{thebibliography}
\end{document}